\documentstyle[amstex,12pt,twoside]{article}            %% AMS-LaTeX
\newlength{\dinwidth}
\newlength{\dinmargin}
\begin{document}
%%%%%%%%%%%%%%%%%%%%%%%%%%%%%%%%%%%%%%%%%%%%%%%%%%%%%%%%%%%%%
\title{Scaling Algebras and Renormalization Group\\ in Algebraic Quantum
  Field Theory.\\ II. Instructive Examples}
\author{ {\sc D. Buchholz} \ and \ 
 {\sc R. Verch}
\\[14pt]
{\normalsize  Institut f\"ur Theoretische Physik,}\\
{\normalsize Universit\"at G\"ottingen,}\\
{\normalsize Bunsenstr. 9,}\\
{\normalsize  D-37073 G\"ottingen, Germany}}
\date{}
\maketitle
%%%%%%%%%Abschnittw. Nummerierung der Gleichungen%%%%%%%%%%%%
\renewcommand{\theequation}{\thesection.\arabic{equation}}
%%%%%%%%%%%%%%%%%%%%%%%%%%%%%%%%%%%%%%%%%%%%%%%%%%%%%%%%%%%%%
%%%%%%%%%Abschnittw. Nummerierung der Theoreme etc.%%%%%%%%%%
\newtheorem{Definition}{Definition}[section]
\newtheorem{Theorem}[Definition]{Theorem}
\newtheorem{Proposition}[Definition]{Proposition}
\newtheorem{Lemma}[Definition]{Lemma}
\newtheorem{Corollary}[Definition]{Corollary}
%%%%%%%%%%%%%%%%%%%%%%%%%%%%%%%%%%%%%%%%%%%%%%%%%%%%%%%%%%%%%%%%%%%%%%
\newsymbol\rest 1316         %% restriction symbol
%%%%%%%%%%%%%%%%%%%%%%%%%%% Newcommands %%%%%%%%%%%%%%%%%%%%%%%%%%%%%%%%
\newcommand{\nin}{\noindent}
%% 
%%%%%%%%%%%%%%%%%%%%%%%%%%%%%%%%%%%%%%%%%%%%%%%%%%%%%%%%%%%%%%%%%%%%%%%%
\newcommand{\CC}{{\mbox{{\small $ \Bbb C$}}}}
\newcommand{\NN}{{\mbox{{\small $ \Bbb N$}}}}
\newcommand{\RR}{{\mbox{{\small $ \Bbb R $}}}}
\newcommand{\II}{{\mbox{{\small $ \Bbb I$}}}}
\newcommand{\ZZ}{{\mbox{{\small $ \Bbb Z$}}}}
\newcommand{\lcrc}{\,{\mbox{{\scriptsize $\circ$}}}\,}
\newcommand{\RRs}{ {\mbox{{\footnotesize $ \Bbb R $}}}}
\newcommand{\RRf}{ {\mbox{{\footnotesize $ \Bbb R $}}}}
\newcommand{\KK}{ {\mbox{{\small $ \Bbb K \,$}} }}
\newcommand{\KKf}{ {\mbox{{\footnotesize $ \Bbb K$ }} }}
\newcommand{\vth}{\vartheta}
\newcommand{\la}{{\lambda}}                     %lambda
\newcommand{\dsx}{d \mbox{\boldmath $x$}}       %measure (spatial) dx
\newcommand{\sx}{\mbox{\footnotesize{\boldmath $x$}}}
                                                %spatial x
\newcommand{\Oo}{{\cal O}}                      %region O
\newcommand{\Op}{{\cal O}'}                     %region O'
\newcommand{\Oi}[1]{{\cal O}_{#1}}              %region O_index
\newcommand{\AO}{{\frak A} ({\cal O})}           %algebra A(O)
\newcommand{\AOp}{{\frak A} ({\cal O}')}         %algebra A(O')
\newcommand{\AOi}[1]{{\frak A} ({\cal O}_{#1})}
                                                %algebra A(O_index)
\newcommand{\Al}{{\cal A }}                     %global algebra A
\newcommand{\fT}{\widetilde{f}}                 %Fourier transf. of f
\newcommand{\BT}{\widetilde{{\cal B}}}          %ball B~
\newcommand{\OT}{\widetilde{{\cal O}}}          %region O~
\newcommand{\AOT}{\widetilde{{\frak A}} ( \widetilde{{\cal O}} )}
                                                %algebra A~(O~)
\newcommand{\AaT}{\widetilde{A}}
\newcommand{\AT}{\widetilde{{\frak A}}}          %algebra A~
\newcommand{\Li}{{\cal L}}                      %left ideal L
\newcommand{\Co}{{\cal C}}                      %algebra C
\newcommand{\Rl}{{R}_{\lambda}}                 %renorm. transformation
\newcommand{\Rml}{{R^{(m)}_{\lambda}}}
\newcommand{\Nl}{N_{\lambda}}                   %renorm. constant
\newcommand{\AlO}{{\frak A}_{\lambda}({\cal O})} %algebra A_lambda(O)
\newcommand{\Alu}{\underline{{\cal A}}}         %global sc. algebra
\newcommand{\AOu}{\underline{{\frak A}}({\cal O})}
                                                 %local sc. algebra
\newcommand{\AOiu}[1]{\underline{{\frak A}} ({\cal O}_{#1})}
                                                 %local sc.algebra index
\newcommand{\Au}{\underline{A}}                  %element of sc. algebra
\newcommand{\Aul}{{\underline{A}}_{\lambda}}     %value of "
\newcommand{\Bul}{{\underline{B}}_{\lambda}}     %value of "
\newcommand{\Abs}[1]{ \left| #1 \right|}
\newcommand{\ABS}[1]{ \left\| #1 \right\|}
\newcommand{\Pg}{{\cal P}_{+}^{\uparrow}}        %Poincare group
\newcommand{\Lx}{ \Lambda , x }                %(Lambda,x)
\newcommand{\ax}{{\alpha}_{{\bf x}}}
\newcommand{\tx}{{\tau}_{{\bf x}}}               
\newcommand{\ay}{{\alpha}_{y}}                   %alpha_y
\newcommand{\az}{{\alpha}_{z}}                   %alpha_z
\newcommand{\af}{{\alpha}_{f}}                   %alpha_f
\newcommand{\ag}{{\alpha}_{g}}                   %alpha_g
\newcommand{\aL}{{\alpha}_{\Lambda}}             %alpha_Lambda
\newcommand{\aLx}{{\alpha}_{\Lambda , x}}        %alpha_{Lambda,x}
\newcommand{\au}{\underline{\alpha}}
\newcommand{\Ux}{U(x)}                           %U(x)
\newcommand{\UL}{U(\Lambda)}                     %U(Lambda)
\newcommand{\ULx}{U(\Lambda , x)}                %U(Lambda,x)
\newcommand{\Uox}{U_{\omega}(x)}                 %U_{omega}(x)
\newcommand{\UoL}{U_{\omega}(\Lambda)}           %U_{omega}(Lambda)
\newcommand{\axu}{{\underline{\alpha}}_x}        %alpha_x sc.
\newcommand{\aLu}{{\underline{\alpha}}_{\Lambda}}
                                                 %alpha_Lambda sc.
\newcommand{\aP}{{\alpha}_{{\cal P}_{+}^{\uparrow}}}
\newcommand{\dmu}{{\underline{\delta}}_{\, \mu}} %delta_mu sc.
\newcommand{\oz}{{\omega}_{0}}                   %vacuum omega_0
\newcommand{\Ho}{{\cal H}_{\omega}}              %H_{omega}
\newcommand{\Hh}{{\cal H}}                       %H
\newcommand{\BH}{{\cal B}({\cal H})}             %B(H)
\newcommand{\Ooo}{{{\Omega}_{\omega}}}           %Omega_omega
\newcommand{\po}{{\pi}_{\omega}}                 %pi_omega
\newcommand{\oziu}{{\underline{\omega}}_{{0}, \iota}}
                                                 %omega_zero_iota sc.
\newcommand{\ou}{{\underline{\omega}}}           %omega sc.
\newcommand{\pu}{{\underline{\pi}}}              %pi sc.
\newcommand{\bra}[1]{\langle {\mbox{$#1$}} |}    %bra
\newcommand{\ket}[1]{| {\mbox{$#1$}} \rangle}    %ket
\newcommand{\ketl}[1]{{|{\mbox{$#1$}} \rangle}_{\lambda}}
                                                 %ket_lambda
\newcommand{\bracket}[2]{\langle {\mbox{$#1$}} | {\mbox{$#2$}} \rangle}
                                                 %bracket
%
\newcommand{\Vek}[1]{
\mbox{ \boldmath $\hspace*{-1mm} #1 \hspace*{-1mm} $ } }
\newcommand{\sVek}[1]{  {\footnotesize
\mbox{ \boldmath $\hspace*{-1mm} #1 \hspace*{-1mm} $ } } }
\newcommand{\FT}[1]{ \widetilde{#1}}
%%%%%%%%More fancy symbols%%%%%%%%%%%%%%%%%%%%%%%%%%%%%%%%%
%
\newcommand{\du}{\underline{\delta}}
\newcommand{\duk}{\underline{\delta}_{\lambda(\kappa)}}
                                              %delta_lambda(kappa) sc.
\newcommand{\dul}{\underline{\delta}_{\lambda}}
\newcommand{\dl}{{\delta}_{\lambda}}
\newcommand{\Hziu}{\underline{\cal H}_{0,\iota}}   %H_0,iota sc.
\newcommand{\Omiu}{\underline{\Omega}_{0,\iota}}   %Omega_0,iota sc.
\newcommand{\pziu}{\underline{\pi}_{0,\iota}}      %pi_0,iota sc.
\newcommand{\tu}{\underline{\tau}}
\newcommand{\aLxu}{\underline{\alpha}_{{\Lambda},x}} %alpha_Lambda,x sc.
\newcommand{\Dziu}{\underline{\Delta}_{0,\iota}}  %Delta_0,i sc.
\newcommand{\Jziu}{\underline{J}_{0,\iota}}       %J_0,i sc.
\newcommand{\Uziu}{\underline{U}_{0,\iota}}       %U_0,i sc.
\newcommand{\Bu}{\underline{B}}
\newcommand{\Cu}{\underline{C}}
\newcommand{\Ak}{A_{\kappa}}
\newcommand{\Bk}{B_{\kappa}}
\newcommand{\auP}{\underline{\alpha}_{{\cal P}_{+}^{\uparrow}}}
\newcommand{\bT}{\widetilde{B}}
\newcommand{\rii}{\rho^{(\iota)}}
\newcommand{\ril}{\rho^{(\iota)}_{\lambda}}
\newcommand{\Hu}{\underline{\cal H}}
\newcommand{\Ou}{\underline{\Omega}}
%% Additional (Update) %%%
\newcommand{\pul}{\underline{\pi}_{\lambda}}
\newcommand{\lk}{\lambda_{\kappa}}
\newcommand{\Aoi}{{\cal A}_{0,\iota}}
\newcommand{\Coi}{C_{0,\iota}}
\newcommand{\aoi}{\alpha^{(0,\iota)}}
\newcommand{\sgl}{\sigma_{\lambda}}
\newcommand{\smu}{\underline{\sigma}_{\mu}}
\newcommand{\sglu}{\underline{\sigma}_{\lambda}}
\newcommand{\olu}{\underline{\omega}_{\lambda}}
\newcommand{\ooi}{\omega_{0,\iota}}
\newcommand{\lkk}{\lambda_{k}}
\newcommand{\oim}{0,\iota(\mu)}
\newcommand{\doim}{\delta_{\mu}^{(0,\iota)}}
\newcommand{\toim}{\tau_{\mu}^{(0,\iota)}}
\newcommand{\Rr}{{\cal R}}
\newcommand{\Wp}{{\cal W}_{+}}
\newcommand{\Wm}{{\cal W}_{-}}
\newcommand{\Uoi}{U_{0,\iota}}
\newcommand{\Omoi}{\Omega_{0,\iota}}
\newcommand{\xu}{\underline{\chi}}
\newcommand{\Iu}{\underline{I}}
\newcommand{\Aa}{{\cal A}}
\newcommand{\Aab}{\bar{\cal A}}
\newcommand{\Lg}{{\cal L}^{\uparrow}_{+}}
\newcommand{\Cin}{C^{\infty}}
\newcommand{\Coin}{C_{0}^{\infty}}
\newcommand{\Cc}{{\cal C}}
\newcommand{\Nn}{{\cal N}}
%%%%%%%%%%%%%%%%%%%%%%%%%%%%%%%%%%%%%%%%%%%%%%%%%%%%%%%%%%%%%%%%%%%%%%%
%%%%%%%%%%  New symbols for paper  2 %%%%%%%%%%%%%%%%%%%%%%%%%%%%%%%%%%
%%%%%%%%%%%%%%%%%%%%%%%%%%%%%%%%%%%%%%%%%%%%%%%%%%%%%%%%%%%%%%%%%%%%%%%
\newcommand{\vx}{{\bf x}}
\newcommand{\vy}{{\bf y}}
\newcommand{\vp}{{\bf p}}
\newcommand{\amx}{\alpha^{(m)}_x}
\newcommand{\amLx}{\alpha^{(m)}_{\Lambda,x}}
\newcommand{\almx}{\alpha^{(\la m)}_x}
\newcommand{\tmx}{\tau^{(m)}_x}
%% { {\mbox{\boldmath $x$}} }
\newcommand{\vxs}{{\bf x}}
%% {{\mbox{{\footnotesize {\boldmath $x$}}}}}
\newcommand{\vO}{{\bf O}}
%% { {\mbox{\boldmath $O$}} }
\newcommand{\vOs}{{\bf O}}
%% {{\mbox{{\footnotesize {\boldmath $O$}}}}}
\newcommand{\dO}{{\cal O}_{\bf O}}
%% { {\cal O}_{{\mbox{{\footnotesize {\boldmath $O$}}}}}}
\newcommand{\mH}{ \hat{\cal H}_m}
\newcommand{\mvac}{ \hat{\Omega}_m}
\newcommand{\mU}{ \hat{U}_m}
\newcommand{\malz}{\alpha^{(m)}}
\newcommand{\amt}{\alpha^{(m)}_t}
\newcommand{\tmt}{\tau^{(m)}_t}
\newcommand{\zH}{ \hat{\cal H}_0}
\newcommand{\zvac}{\hat{\Omega}_0}
\newcommand{\zalf}{\hat{\alpha}^{(0)}}
\newcommand{\zsig}{\hat{\sigma}}
\newcommand{\Feld}{\hat{\Phi}_m}
\newcommand{\Felz}{\hat{\Phi}_0}
\newcommand{\Weyl}{{\frak W}}
\newcommand{\Ll}{{\cal L}}
\newcommand{\mg}{\gamma_m}
\newcommand{\mvst}{\hat{\omega}^{(m)}}
\newcommand{\mvsz}{\omega^{(m)}}
\newcommand{\mmp}{\mu_{m,{\bf p}}}
\newcommand{\vk}{{\bf k}}
%% {\mbox{\boldmath $k$}}
\newcommand{\vks}{{\bf k}}
%% {\mbox{\footnotesize {\boldmath $k$}}}
\newcommand{\va}{{\bf a}}
%% {\mbox{\boldmath $a$}}
\newcommand{\vas}{{\bf a}}
%% {\mbox{{\footnotesize {\boldmath $a$}}}}
\newcommand{\mok}{\mu_{m,{\bf k}}}
\newcommand{\mHz}{{\cal H}_m}
\newcommand{\mpz}{\pi_m}
\newcommand{\mvaz}{\Omega_m}
\newcommand{\mdom}{{\cal D}_m}
\newcommand{\AAm}{{\cal A}^{(m)}}
\newcommand{\Aam}{{\cal A}^{(m)}}
\newcommand{\mBB}{{\frak B}^{(m)}}
\newcommand{\umBB}{\underline{\frak B}^{(m)}}
\newcommand{\umalf}{\underline{\alpha}^{(m)}}
\newcommand{\mvoi}{\omega^{(m)}_{0,\iota}}
\newcommand{\maoi}{\alpha^{(m;0,\iota)}}
\newcommand{\umvs}{\underline{\hat{\omega}}^{(m)}}
\newcommand{\uAAm}{\underline{\cal A}^{(m)}}
\newcommand{\oBB}{{\frak B}^{(0)}}
\newcommand{\mWeyl}{{\frak W}^{(m)}}
\newcommand{\umWeyl}{\underline{\frak W}^{(m)}}
\newcommand{\umalz}{\underline{\alpha}^{(m)}}
\newcommand{\Wu}{\underline{W}}
\newcommand{\mAoi}{{\cal A}^{(m)}_{0,\iota}}
\newcommand{\mvzi}{\omega^{(m)}_{0,\iota}}
\newcommand{\hHoi}{\hat{\cal H}_{0,\iota}}
\newcommand{\hpoi}{\hat{\pi}_{0,\iota}}
\newcommand{\hOoi}{\hat{\Omega}_{0,\iota}}
\newcommand{\poi}{\pi_{0,\iota}}
\newcommand{\Hoi}{{\cal H}_{0,\iota}}
\newcommand{\Ooi}{\Omega_{0,\iota}}
\newcommand{\mazi}{\alpha^{(m;0,\iota)}}
\newcommand{\oip}{\omega_{q,\iota}^{(m)}}
\newcommand{\dml}{\delta^{(m)}_{\lambda}}
\newcommand{\sml}{\sigma^{(m)}_{\lambda}}
\newcommand{\Woi}{W_{0,\iota}}
\newcommand{\DRs}{{\cal D}({\mbox{{\small $ \Bbb R $}}}^s)}
\newcommand{\DR}{{\cal D}({\mbox{{\small $ \Bbb R $}}})}
\renewcommand{\epsilon}{\varepsilon}
\newcommand{\rqi}{\varrho_{q,\iota}}
\newcommand{\Vqn}{V_q^{(n)}}
\newcommand{\Yqn}{Y_q^{(n)}}
\newcommand{\Ome}{{\Omega^{(0)}}}
%
%%%%%%%%%%%%%%%%%%%%%%%%%%%%%%%%%%%%%%%%%%%%%%%%%%%%%%%%%%%%%%%%%%%%%%%%%%
%%%%%%%%%%%%%%%%%%%%%%%%%%%%%%%%%%%%%%%%%%%%%%%%%%%%%%%%%%%%%%%%%%%%%%%%%%
%%%%%%%%%%%%%%%%%%%%%%%%%%%%%%%%%%%%%%%%%%%%%%%%%%%%%%%%%%%%%%%%%%%%%%%%%%
%%%%%%%%%%%
%%%%%%%%%%%%%%%%%%%%%%%%%%%%%%%%%%%%%%%%%%%%%%%%%%%%%%%%%%%%%%%%%%%%%%%
%%%%%%%%%%%%%%%%%%%%%%%%%%%%%%%%%%%%%%%%%%%%%%%%%%%%%%%%%%%%%%%%%%%%%%%%%
\noindent
{\small  {\bf Abstract:}
  The concept of scaling algebra provides a novel framework for the
  general structural analysis and classification of the short distance
  properties of algebras of local observables in relativistic quantum
  field theory. In the present article this method is applied to the
  simple example of massive free field theory in $s = 1,2$ and $3$
  spatial dimensions. Not quite unexpectedly, one obtains
  for $s = 2,3$ in the scaling
  (short distance) limit the algebra of local observables in massless
  free field theory. The case $s =1$ offers, however, some surprises.
  There the
  algebra of observables acquires in the scaling limit a non-trivial
  center and describes charged physical states satisfying Gauss'
  law. The latter result is of relevance for the interpretation of the
  Schwinger model at short distances and illustrates the conceptual
  and computational virtues of the method.}
${}$ \\[14pt]
%%%%%%%%%%%%%%%%%%%%%%%%%%%%%%%%%%%%%%%%%%%%%%%
\section{Introduction}
%%%%%%%%%%%%%%%%%%%%%%%%%%%%%%%%%%%%%%%%%%%%%%%
The structure of local observables in relativistic quantum field
theories at short distances is in many respects of physical
interest. It is of relevance for the interpretation of physical states
at small spacetime scales, the classification of the possible
ultraviolet properties of local observables and the clarification of
the corresponding algebraic structures. One may also hope  that a
general model-independent understanding of this issue will shed 
light on the constructive problems in local quantum physics \cite{Ha}.
 
A promising step towards the solution of these problems has been taken
in \cite{BV}, where the basic ideas of renormalization group theory
have been adapted to the Haag-Kastler framework of relativistic
quantum field theory. It was shown in this analysis that to each
algebra of local observables there exists an associated scaling
algebra on which renormalization group (scaling) transformations act
in a canonical manner. With the help of this device one can define and
classify the scaling limit of any given theory \cite{BV}, analyze the
relation between phase space properties and the nature of the scaling
limit \cite{Bu96a} and introduce concepts for the description of the
particle and symmetry features of a theory at small scales \cite{Bu96}.

It is the aim of the present article to illustrate the computational
aspects of this method by applying it to the simple case of 
the theory of a massive
free scalar field in $s$ spatial dimensions. Following is a very brief
outline of our results; a more detailed summary will be given at the
end of this introduction.  

 It turns out that, in the
cases $s=2,3$, one obtains in the scaling limit the algebra of
observables in the corresponding massless free field theories. Thus,
according to the classification in \cite{BV}, these theories have a
unique quantum scaling limit.
Thinking of the conventional field-theoretic approach to the
renormalization group, this result may not be unexpected. Nevertheless
it is of interest since it illustrates  the basic message of
\cite{BV} that for the short distance analysis one need not exhibit
specific renormalization group transformations. It is sufficient to
identify the observables at different scales by considering
operator-functions of a scaling parameter which have a few general
properties and which exist in abundance. This somewhat abstract
approach has the virtue of being model independent, but it is still
sufficiently concrete in order to carry out explicit computations.

In the case of free field theory in $s=1$ spatial dimensions, where
the conventional field theoretic approach to the short distance
analysis is hampered by infrared problems, the method of the scaling
algebra reveals its full strength. There it turns out that the algebra
of observables in the scaling limit is a (central) extension of the
algebra generated by the massless free field in exponentiated Weyl
form.
The presence of a center shows that, for $s=1$, the vacuum states
appearing in the scaling limit can be mixed, in contrast to theories
in higher dimensions, where these states are always pure \cite{BV}.

More interestingly, the present method allows one to exhibit in the
scaling limit physical states carrying a (global) gauge charge in the
sense of \cite{BF} for which Gauss' law holds. This result provides a
marked illustration of the fact that the charge structure of a theory
may differ substantially from that of its scaling limit. It is in
particular of relevance for the interpretation of the Schwinger model
where the algebra of the local (gauge-invariant) observables is known
to be isomorphic to that generated by a free massive field
\cite{LoSw}. Thus, whereas  this theory does not have any charged
superselection sectors at finite scales, there appear physical states
in the scaling limit carrying an ``electric charge''. The presence of
these states may be interpreted as a manifestation of
``partons'' in the Schwinger model, i.e.\ observable particle-like structures
appearing at small spacetime scales which have no counterpart at large
scales. For a more detailed discussion of this issue and its relation
to the notion of confinement, cf.\ \cite{Bu96}.

Hence, even though the models underlying our present investigation are
rather trivial, the results nicely illustrate and exemplify various
points in the abstract analysis carried out in \cite{BV}.

For the convenience of the reader we recall in the remainder of this
introduction various notions and results from \cite{BV} and establish
our notation. The method of the scaling algebra relies  on the
following fundamental properties of any physically acceptable theory
\cite{Ha}.\\[6pt]
1. {(Locality)} The observables of the theory generate a net of
local algebras over $(1+s)$-dimensional Minkowski space $\RR^{1+s}$,
i.e.\ an inclusion preserving map
\begin{equation}
\Oo \to \Aa(\Oo)
\end{equation}
from the set of open double cones $\Oo \subset \RR^{1+s}$ to unital
$C^*$-algebras $\Aa(\Oo)$. The algebra generated by all local algebras
$\Aa(\Oo)$  (as a $C^*$-inductive limit) is denoted by $\Aa$. The net
is supposed to satisfy the principle of locality (Einstein causality),
i.e.\ all pairs of operators which are assigned to spacelike separated
double cones commute. \\[6pt]
2. {(Covariance)} The Poincar\'e group $\Pg$ is represented by
automorphisms of the net. Thus for each $(\Lx) \in \Pg$ there is an
automorphism $\aLx \in {\rm Aut}\,\Aa$ such that, in an obvious
notation,
\begin{equation}
 \aLx(\Aa(\Oo)) = \Aa(\Lambda \Oo + x)
\end{equation}
for any double cone $\Oo$. In \cite{BV} this fundamental
postulate was amended 
by the condition that for each
$A\in\Aa$ the function $(\Lx) \mapsto \aLx(A)$ is strongly continuous.
In the present analysis we require continuity only with respect to the
translations.
\\[6pt]
3. {(States)} The physical states are described by positive,
linear and normalized functionals $\omega$ on $\Aa$. By the
GNS-construction, any state $\omega$ gives rise to a representation
$\pi_{\omega}$ of $\Aa$ on a Hilbert space $\Hh_{\omega}$, and there
exists a unit vector $\Omega_{\omega} \in \Hh_{\omega}$ such that
\begin{equation}
\omega(A) = \langle
\Omega_{\omega},\pi_{\omega}(A)\Omega_{\omega}\rangle\,,\quad a \in
\Aa\,. 
\end{equation}
States describing the vacuum are distinguished by the fact that, on
the corresponding Hilbert space $\Hh_{\omega}$, there is a continuous
unitary representation $U_{\omega}(\Lx)$ of $\Pg$ which leaves the
unit vector $\Omega_{\omega}$ invariant, satisfies the relativistic
spectrum condition (positivity of the energy) and implements the
action of $\Pg$ on the observables,
\begin{equation} 
 U_{\omega}(\Lx)\pi_{\omega}(A)U_{\omega}(\Lx)^{-1} = \pi_{\omega}
(\aLx(A))\,, \quad A \in \Aa\,.
\end{equation}
Any state of physical interest is assumed to be locally normal to the
vacuum state (i.e.\ its restriction to any local algebra can be
represented by a vector in the Hilbert space of the vacuum
representation,
cf.\ \cite[Sec.\ V.2]{Ha}).\\[6pt]

We amend these physically well-motivated assumptions by a condition of
a more technical nature in order to simplify the subsequent
discussions. Namely we assume that the local algebras are continuous
from the outside,
\begin{equation}
\Aa(\Oo) = \bigcap_{\Oo_1 \supset \overline{\Oo}} \Aa(\Oo_1)\,,
\end{equation}
where $\overline{\Oo}$ denotes the closure of $\Oo$. If a given net 
$\Oo \to \Aa(\Oo)$ fails to comply with this additional condition, one
can always proceed to a corresponding regularized net
$\Oo \to \Aa_{\rm reg}(\Oo): = \bigcap_{\Oo_1 \supset \overline{\Oo}}
\Aa(\Oo_1)$
which has all the desired properties, as is easily checked. In view of
this simple fact we may assume without essential loss of generality
that all local nets appearing in our analysis are continuous in the
sense of relation (1.5).

Within this setting the short distance analysis is carried out as
follows. One first proceeds from the given net and automorphisms
$(\Aa,\alpha)$ at spacetime scale $\la =1$ (in appropriate units) to the
corresponding nets $(\Aa_{\la},\alpha^{(\la)})$ describing the theory
at arbitrary scale $\la \in \RR^+$. This is easily done by
setting for given $\la > 0$
\begin{equation}
\Aa_{\la}(\Oo):= \Aa(\la \Oo)\,,\quad \ \ \alpha^{(\la)}_{\Lx}:=
\alpha_{\Lambda,\la x}\,.
\end{equation}
The identification of observables at different scales can then be
accomplished by considering functions $\Au$ of the scaling parameter
$\la > 0$
whose values $\Aul$  are to be interpreted as
observables in the nets $(\Aa_{\la},\alpha^{(\la)})$, $\la \in \RR^+$.
Clearly, any such function $\Au$ establishes some relation between observables
at different scales.

With this simple idea in mind one is led to the concept of scaling
algebra $\Alu$ which consists of certain specific functions
$\Au : \RR^+ \to \Aa$ with properties described below. The algebraic
operations in $\Alu$ are pointwise defined by the corresponding
operations in $\Aa_{\la}$, $\la \in \RR^+$,
 and there is a $C^*$-norm on $\Alu$ given by
\begin{equation}
||\,\Au\,|| := \sup_{\la > 0}\,||\,\Aul\,||\,.
\end{equation}
The local structure of $\Aa$ is lifted to $\Alu$ by setting
\begin{equation}
\Alu(\Oo) := \{ \Au : \Aul \in \Aa_{\la}(\Oo)\,,\ \ \  \la \in
\RR^+\}\,.
\end{equation}
Hence $\Oo \to \Alu(\Oo)$ is a local net over $\RR^{1+s}$ and $\Alu$
is defined as its $C^*$-inductive limit. One can also lift the action
of the Poincar\'e group in the underlying theory to a corresponding
action of automorphisms $\au_{\Lx}$ on $\Alu$, which are given by
\begin{equation}
(\au_{\Lx} (\Au) )_{\la} := \alpha_{\Lambda,\la x}(\Aul)\,.
\end{equation}
It is crucial to demand that $\Alu$ consists only of elements on which
the translations act strongly continuously, i.e.\ for each $\Au \in
\Alu$ there holds 
\begin{equation}
||\,\au_x(\Au) - \Au\,|| \to 0 \quad {\rm as} \quad x \to 0\,.
\end{equation}
Heuristically speaking, the latter constraint amounts to the condition
that, for given $\Au$, the operators $\Aul$ occupy, for all values of
$\la$, certain regions in ``phase space'' with a fixed volume
\cite{BV}. Hence, whereas the scale of spacetime changes along the
graph of $\Au$, the scale $\hbar$ of action is kept fixed. In
\cite{BV} it was assumed that also the Lorentz transformations act in
a strongly continuous manner on $\Alu$; yet in order to simplify the
discussion we do not impose this stronger condition here and consider
the somewhat larger scaling algebra $\Alu$ consisting of all functions
satisfying the preceding, weaker conditions. It should be emphasized
that this algebra is fixed by these conditions once a local net
$(\Aa,\alpha)$ is given.

The structure of the physical states $\omega$ in the underlying theory
at small spacetime scales can now be analyzed as follows. Given
$\omega$ one defines a lift of this state to the scaling algebra at
scale $\la \in \RR^+$ by setting
\begin{equation}
\olu(\Au):= \omega(\Aul)\,, \quad \Au \in \Alu\,.
\end{equation}
Let $(\pi_{\la},\Hh_{\la})$ be the GNS-representation of $\Alu$ which
is fixed by $\olu$. Then one considers the net
\begin{equation}
 \Oo \to \Alu(\Oo)/{\rm ker}\,\pi_{\la}\,,\quad\au^{(\la)}_{\Lx}\,,
\end{equation}
where ker means ``kernel'' and $\au^{(\la)}_{\Lx}$ is the induced action of
the Poincar\'e-transfor\-mations $\aLxu$
on this quotient. This net is
isomorphic to the theory $(\Aa_{\la},\alpha^{(\la)})$ at scale $\la$
\cite{BV}.

 We recall in this context that two local nets
$(\Aa_{a},\alpha^{(a)})$ and $(\Aa_{b},\alpha^{(b)})$ are said to be
isomorphic if there exists an isomorphism $\phi : \Aa_{a} \to
\Aa_{b}$ which preserves locality, $\phi(\Aa_{a}(\Oo)) =
\Aa_{b}(\Oo)$ for each double cone $\Oo$, and intertwines the action
of the Poincar\'e-transformations, $\phi \lcrc \alpha^{(a)}_{\Lx} =
\alpha^{(b)}_{\Lx}\lcrc \phi$; isomorphic nets describe the same
physics.

With these preparations one is led to the following canonical
definition of the scaling limit of a theory. One first considers the
set $SL(\omega)$ of limit points (in the weak-*-topology) of the net
of states $\{\olu\}_{\la >0}$ for $\la \to 0$. This set of states is
always non-empty by standard compactness arguments. We denote the
elements of $SL(\omega)$ by $\ooi$, where $\iota$ is an element of
some index set, and recall that $\ooi \in SL(\omega)$ means that there
exists some directed set $\KK$ (depending on $\iota$) such that, for
some net of scaling parameters $\lk$, $\kappa \in \KK$, which
converges to zero, one has
\begin{equation}
 \lim_{\kappa}\, \ou_{\lk}(\Au) = \ooi(\Au) \,,\quad \Au \in \Alu\,.
\end{equation}
Thus, roughly speaking,  the sequence of states $\olu$ need
not converge for $\la \to 0$, but there exist always convergent
subsequences.
The following general facts about the scaling limit states $\ooi$ have
been established in \cite{BV}.
\begin{itemize}
  \item[1.] $SL(\omega)$ does not depend on the chosen physical
    (locally normal) state $\omega$.
  \item[2.] Each $\ooi \in SL(\omega)$ is a vacuum state on
    $\Alu$. (In the present, more general setting one has, however, no control
    on the continuity properties of the Lorentz transformations.) 
\end{itemize}
With this information it is clear how to define, in analogy to the
case of finite scales, theories associated with the scaling limit $\la
\to 0$. Picking $\ooi \in SL(\omega)$ one proceeds to its
GNS-representation $(\poi,\Hoi)$ and defines the net and automorphisms
\begin{equation}
\Oo \to \Aoi(\Oo) := \Alu(\Oo)/{\rm ker}\, \poi\,,\quad \aoi_{\Lx}\,,
\end{equation}
where $\aoi_{\Lx}$ denotes the induced action of $\au_{\Lx}$
on the quotient net.
This net has the same general properties as the underlying theory
(possibly apart from outer continuity; in that case we pass to the
regularized net without further mentioning).

We mention as an aside that in our computations we will frequently
make use of the general fact that the abstract net defined
 in (1.14) is isomorphic to the
concrete net of $C^*$-algebras and automorphisms given by 
\begin{equation}
\Oo \to \poi(\Alu(\Oo))\,,\quad {\rm Ad}\,\Uoi(\Lx)\,,
\end{equation}
where $\Uoi(\Lx)$ are the unitaries representing the Poincar\'e group
on $\Hoi$.

Given these results, there arises the interesting problem of whether
the nets $(\Aoi,\aoi)$ depend on the choice of the state $\ooi \in
SL(\omega)$. As was discussed in \cite{BV}, there are the following
possibilities.
\begin{itemize}
\item[1.] All nets $(\Aoi,\aoi)$ are isomorphic to the trivial net
  $(\CC\, 1, \mbox{id})$, where id denotes the trivial automorphism
  (``classical scaling limit'').
\item[2.] All nets $(\Aoi,\aoi)$ are isomorphic and non-trivial
  (``unique quantum scaling limit''). If the respective isomorphisms
  connect also the vacuum states $\ooi$, one has ``a unique vacuum
  structure in the scaling limit''.
\item[3.] Not all of the nets $(\Aoi,\aoi)$ are isomorphic (``degenerate
  scaling limit'').
\end{itemize}
Which case is at hand depends of course on the underlying theory. As
was argued in \cite{BV}, case 2 may be expected to hold in many theories of
physical interest. In spite of the existence of an abundance of scaling
limit states $\ooi$, which may be attributed to the fact that the
scaling algebra $\Alu$ contains  the orbits of local observables
under arbitrary renormalization group transformations \cite{BV}, these
theories have a well-defined and non-trivial scaling limit. Phrased
differently, in this generic case it does not matter which
renormalization group transformation one uses in order to determine
the scaling limit, all transformations yield the same result.

 The same remark applies to theories leading to case 1, for which the
 scaling limit is trivial. It is only for theories corresponding to
 case 3 that the short distance structure cannot be described by a
 single net. For a tentative physical interpretation of these cases
 see \cite{BV}. Examples of  nets with such remarkable short distance
 properties will be presented elsewhere
 \cite{BuLu}.

In the present investigation we will apply this scheme to the nets
$(\Aa^{(m)},\alpha^{(m)})$ generated by the free field of mass $m$ in
$s$ spatial dimensions. These nets will be introduced in Sec.\ 2 as
concrete operator algebras in some ``standard representation'' which
differs from the familiar Fock representation but will be convenient
for the present analysis since it accommodates the free nets of
arbitrary mass in a transparent manner. In Sec.\ 3 we will analyze the
structure of the scaling limit nets $(\mAoi,\maoi)$ in the cases $s=2,3$
and show that they are isomorphic to the net $(\Aa^{(0)},\alpha^{(0)})$
generated by the free massless field and that the isomorphisms connect
the respective vacuum states. Hence, in this specific sense these
theories have a unique quantum scaling limit and vacuum structure in
this limit.

The case $s=1$ is treated in Sec.\ 4, where we show that, as already
mentioned, the algebras $\mAoi$ have a non-trivial center and hence
the vacuum states $\mvzi$ are mixed. We also exhibit physical states
$\oip$ on the nets $(\mAoi,\maoi)$ which carry some charge $q$ and
coincide with the vacuum state $\mvzi$ on the observables in the right
and left spacelike complement of some sufficiently large double
cone. But they are disjoint from the vacuum on the algebra of
observables of the full spacelike complement of any double cone, no
matter how large. Therefore one can determine  the charge
of these states in the spacelike complement of any bounded region and
this fact may be regarded as an algebraic version of Gauss' law. The
article concludes with a list of open problems which are
outlined in Sec.\ 5.

\section{Standard representations of free fields}
\setcounter{equation}{0}
%%%%%%%%%%
We deal in the present article with the theories of the free scalar
field of arbitrary mass. It is therefore convenient to employ the
formulation of free field theory based on the time zero fields and
their canonically conjugate momenta since they do not depend on the mass.

We introduce the fields and conjugate momenta in exponentiated form by
considering the unitary operators $W(f)$, where $f$ is any element of 
$\DRs$, the space of complex valued test-functions with compact
support in the configuration space $\RR^s$. The canonical commutation
relations then turn into the Weyl relations
\begin{equation}
W(f)W(g) = {\rm e}^{- \frac{i}{2}\sigma(f,g)}W(f+g)\,, \quad f,g \in
\DRs\,, 
\end{equation}
where the symplectic form $\sigma$ is given by
\begin{equation}
\sigma(f,g) := {\rm Im}\, \int d^sx\, \overline{f(\vx)} g(\vx)\,.
\end{equation}
The *-algebra generated by all Weyl operators is denoted by $\Weyl$.
On $\Weyl$ we introduce various automorphisms of geometrical
significance. The action of the spatial translations $\RR^s$ on the
Weyl operators is given by 
\begin{equation}
\ax(W(f)) := W(\tx f)\,, \quad \vx \in \RR^s\,,
\end{equation}
where $(\tx f)(\vy) := f(\vx - \vy)$. For given mass $m \ge 0$, we
define corresponding time translations by setting
\begin{equation}
\amt(W(f)) := W(\tmt f)\,, \quad t \in \RR\,.
\end{equation}
Here, $(\tmt f)(\vx)$ is the unique solution of the Klein-Gordon
equation of mass $m$ with initial data ${\rm Re}f + i{\rm Im}f$,
$(\Delta - m^2){\rm Im}f + i{\rm Re}f$.  More
explicitly, 
\begin{eqnarray}
(\tmt f) & := & ( \cos(t \mu_m) + i \mu_m^{-1}\sin(t \mu_m))
{\rm Re} f \\
& + &  i \left( \cos(t \mu_m) + i \mu_m\sin(t \mu_m)^{{}}\right) {\rm
  Im}f\,,
\nonumber
\end{eqnarray}
where $\mu_m$ acts in momentum space according to $\widetilde{
(\mu_m f)}(\vp) := \sqrt{\vp^2 + m^2}\tilde{f}(\vp)$. (The
tilde denotes, as usual, the Fourier-transform.) Because of the
propagation properties of the solutions of the Klein-Gordon equation,
$(\tmt f)$ has support in a ball of radius $r + |t|$ if $f$ has support in
a ball of radius $r$; hence $\DRs$ is stable under the action of
$\tmt$.

It is apparent that the automorphisms $\ax$ and $\amt$ commute for
arbitrary $m \ge 0$. In contrast, the time translations corresponding
to different values of $m$ do not commute. In order to simplify
notation we put $\amx := \amt \lcrc \ax$ and similarly $\tmx := \tmt
\lcrc \tx$ for $x = (t,\vx) \in \RR^{1 +s}$.

In a similar way one can introduce a mass dependent action
$\malz_{\Lambda}$ of the
Lorentz-transformations on $\Weyl$, but we
dispense with giving explicit formulas. We will also need
the action of length scale transformations (dilations) on $\Weyl$.
They are fixed by setting
\begin{equation}
\sgl(W(f)) := W(\dl f)\,, \quad \la > 0,
\end{equation}
where
\begin{equation}
(\dl f)(\vx) := \la^{- \frac{s+1}{2}}({\rm Re}f)(\la^{-1}\vx)
 + i \la^{- \frac{s -1}{2}}({\rm Im} f)(\la^{-1}\vx)\,.
\end{equation}
It is straightforward to establish the following relation between the
Poincar\'e-trans\-for\-mations and dilations:
\begin{equation}
 \sgl \lcrc 
\alpha^{(\la m)}_{\Lx} = \alpha^{(m)}_{\Lambda,\la x} \lcrc \sgl\,,
 \ \ \la > 0\,.
\end{equation}
Next we introduce the vacuum states on $\Weyl$ corresponding to the
different time evolutions. Given $m \ge 0$, we put
\begin{equation}
 \mvsz(W(f)) := {\rm e}^{- \frac{1}{2}||\,f\,||_m^2}\,, \quad f \in
 \DRs\,,
\end{equation}
where (excluding the singular case $s = 1$, $m = 0$) 
\begin{equation}
||\,f \,||^2_m := 2^{-1}\,
\int d^sp \, |\, \mmp^{-1/2}(\widetilde{{\rm Re}f})(\vp)
 + i \mmp^{1/2} (\widetilde{{\rm Im}f})(\vp)\,| ^2
\end{equation}
and $\mmp := \sqrt{\vp^2 + m^2}$. The extension of $\mvsz$ to $\Weyl$
is fixed by linearity and describes the vacuum state in the theory of
mass $m$.
There holds in particular $\mvsz \lcrc \alpha^{(m)}_{\Lx} = \mvsz$ and 
$\mvsz \lcrc \sgl = \omega^{(\la m)}$.

The present analysis is greatly simplified by the following result due
to Eckmann and Fr\"ohlich \cite{EF}. In its formulation there enter
the sub-algebras $\Weyl(G)$ of $\Weyl$ which are assigned to the
regions $G \subset \RR^s$ and are generated by all Weyl operators
$W(g)$
where $g$ are test-functions with ${\rm supp}\, g \subset G$.
\begin{Proposition}
Let $s = 2$ or $3$ and let $G \subset \RR^s$ be bounded. Then the
restricted (partial) states $\mvsz \rest \Weyl(G)$, $m \ge 0$, are
normal with respect to each other. If $s =1$ this statement holds for
$m > 0$.
\end{Proposition}
This result and the local action of the spacetime transformations allow
one to describe and analyze the nets of local von Neumann algebras
generated by the various free fields of different mass in a fixed 
``standard representation'' of $\Weyl$ which is induced by any one of
the vacuum states. In the subsequent section we treat the cases 
$s = 2,3$ and take as a standard state the $m =0$ vacuum, whereas in
our discussion of the case  $s=1$, we fix some
vacuum state with $m  > 0$. In order to simplify notation, we
use the symbol $W(f)$ also for the concrete Weyl operators in the
GNS-representation of $\Weyl$ induced by the chosen standard state.

Within the chosen standard representation we can define the net of local von
Neumann algebras on Minkowski space corresponding to the theory of
mass $m$ as follows. Given any double cone $\Oo_0 \subset \RR^{1+s}$
with base $G$
in the time $t =0$ plane and any Poincar\'e transformation $\Lx$,
we set
\begin{equation}
 \Rr^{(m)}(\Lambda\Oo_0 + x) := \{\amLx(W(g)) : {\rm supp}\,g \subset G
 \}''\,,
\end{equation}
where the prime denotes the commutant.
 In this way we
obtain a local net $\Oo \to \Rr^{(m)}(\Oo)$ of von Neumann algebras on
the underlying Hilbert space which,
by the result of Eckmann and Fr\"ohlich, is isomorphic to the net generated
by the free field of mass $m$  on the Fock space corresponding to
$\mvsz$. Moreover, the automorphisms $\malz_{\Lx}$ extend to the local von
Neumann algebras $\Rr^{(m)}(\Oo)$ and act covariantly on the net, i.e.
\begin{equation}
 \amLx(\Rr^{(m)}(\Oo)) = \Rr^{(m)}(\Lambda\Oo+ x)\,.
\end{equation}
Note, however, that for $m$ different from the mass of the chosen
standard state the time translations $\amt$ are  not 
unitarily implemented in the underlying Hilbert space. An analogous
statement holds for the Lorentz boosts.

Let us briefly indicate the advantage of the present standard
representation of the various local nets which is only locally normal
with respect to the familiar Fock representations.
What we gain is the equality of the local algebras $\Rr^{(m)}(\Oo_0)$
for arbitrary $m$ and any double cone $\Oo_0$ with base in the time 
$t =0$ plane, cf.\ relation (2.11). As a consequence there holds for
arbitrary double cones $\Oo$ and masses $m_1$,$m_2$
\begin{equation}
\Rr^{(m_1)}(\Oo)  \subset \Rr^{(m_2)}(\Oo_0)
\end{equation}
whenever $\Oo_0$ is some double cone with base at $t =0$ which
contains $\Oo$. Moreover, choosing $\omega^{(0)}$ as the standard
state in the cases $s =2,3$, we are able to  use
in the analysis of the massive theories the invariance of
$\omega^{(0)}$ under the
 dilations $\sgl$, and their covariant action on the massless net,
\begin{equation} \sgl(\Rr^{(0)}(\Oo)) = \Rr^{(0)}(\la \Oo)\,. 
\end{equation}
 Since the scaling limit of a theory does not depend on the
choice of a locally normal state \cite{BV}, the present setting proves
to be most convenient.

As was mentioned in the Introduction, the short distance analysis of a
local net of von Neumann algebras requires the passage to a
corresponding subnet of $C^*$-algebras consisting of operators which
transform strongly continuously under the action of Poincar\'e
transformations or, more generally, spacetime translations. We
restrict attention here to the latter case and consider for fixed $m$
the weakly dense subnet of $\Oo \to \Rr^{(m)}(\Oo)$ given by
\begin{equation}
\Oo \to \Aam(\Oo) := \{ A \in \Rr^{(m)}(\Oo) : \lim_{x \to 0}\,
 ||\,\amx(A) - A \,|| = 0 \}\,.
\end{equation}
This net still transforms covariantly under the Poincar\'e transformations
 $\amLx$
and,
in the case $m = 0$, also under dilations $\sgl$. Its $C^*$-inductive
limit is denoted by $\Aam$ and the various vacuum states extend to
this algebra by local normality under the conditions stated in
Proposition 2.1.
We also note that the algebras $\Aa^{(m)}(\Oo)$ are continuous from
 the outside,
\begin{equation} \Aa^{(m)}(\Oo) = \bigcap_{\Oo_1 \supset \overline{\Oo}}
\Aa^{(m)}(\Oo_1) \,, 
\end{equation} 
as a consequence of the outer continuity of the von Neumann algebras
$\Rr^{(m)}(\Oo)$.
 We emphasize, however, that because of the continuity
requirement in (2.15) it is no longer true that the algebras $\Aam(\Oo_0)$
coincide for different $m$ and fixed double cones $\Oo_0$ based at time
$t=0$, in contrast to their weak closures.
%%%%%%%%%%%%%%%%%%%%%%%%%%%%%%%%%%%%%%%%%%%%%%%%%%%%%%%%%%%%%%%
\section{Computation of the scaling limit for $s= 2,3$}
\setcounter{equation}{0}
%%%%%%%%%%%%%%%%%%%%%%%%%%%%%%%%%%%%%%%%%%%%%%%%%%%%%%%%%%%%%%%
In the present section our objective is to prove the following result
which provides full information about the scaling limit theories
of the free scalar fields of any mass in three- and four-dimensional
Minkowski-spacetime.
\begin{Theorem}
Let $s =2,3$, $m \ge 0$, and let 
$\mvoi$ be any scaling limit state of the theory
$(\Aa^{(m)},\alpha^{(m)},
\omega^{(m)})$ of a
 free scalar field of mass $m$ in
$(1 +s)$-dimensional Minkowski-spacetime. Then the associated scaling
limit theory $(\Aoi^{(m)},\maoi,\mvoi)$ is net-isomorphic
to the theory $(\Aa^{(0)},\alpha^{(0)},
\omega^{(0)})$ of the
massless free scalar field in the same spacetime dimension,
and the corresponding net-isomorphism connects $\mvoi$ and
$\omega^{(0)}$. 
\end{Theorem}
{\it Remark.} This result implies that,
 according to the classification in \cite{BV}, these free field
 theories have a unique quantum scaling limit with a unique vacuum
 structure.
 A similar theorem holds for the scaling limit theories of the local
 nets
if one imposes the continuity requirements (2.15) and (1.10) for the whole
 Poincar\'e group.
\\[10pt]
The proof of this result proceeds in several steps.
To begin with
we recall (cf.\ the discussion in Sec.\ 2) that we are working in the
standard representation of $\Weyl$ which is induced by
 the mass zero vacuum state $\omega^{(0)}$.
 Now let
$\mvzi$ be a scaling limit state of $\omega^{(m)}$, so that
\begin{equation}
\mvzi(\Au) = \lim_{\kappa}\,\underline{\omega}^{(m)}_{\lk}(\Au)\,,
\quad \Au \in \uAAm\,,
\end{equation}
for a suitable subnet $\lk$, $\kappa \in \KK$, of positive
real numbers converging to 0. We denote the GNS-representation of
$\mvzi$ by $(\poi,\Hoi,\Ooi)$. It is our aim to show that the
required net-isomorphism $\phi$ is obtained by assigning to
$\poi(\Au)$, $\Au \in \uAAm(\Oo)$, the operators
\begin{equation}
\phi(\poi(\Au)) := w-\lim_{\kappa}\,\sigma_{\lk}^{-1}(\Au_{\lk}) \,.
\end{equation}
We must demonstrate that the assignment (3.2) is well-defined and
has the properties needed of a net-isomorphism. To begin with, we
list some useful auxiliary results.
\begin{Lemma} ${}$\\[4pt]
(a) \quad $\lim_{\la \to 0}\,||\,(\omega^{(\la m)} - \omega^{(0)})
\rest \Rr^{(0)}(\Oo)\,|| = 0$ for any double cone $\Oo$.
\\[4pt]
(b) \quad
Let $h \in {\cal D}(\RR^{1+s})$ and
 $f \in \DRs$, and consider the function
 $\Wu: \RR^+ \to \Aa^{(m)}$ given by
\begin{equation}
\Wu_{\la} := \int d^{1+s}x\,h(x)\,\malz_{\la x} \lcrc \sigma_{\la}
(W(f))\,,
\quad \la > 0\,,
\end{equation}
where the integral is to be understood in the weak sense.
Then $\Wu \in \uAAm(\Oo_0)$ for some double cone $\Oo_0$ based on the
 time $t =0$ plane. Moreover,
\begin{equation}
\lim_{\la \to 0}\, \sigma_{\la}^{-1}(\Wu_{\la}) =
 \int d^{1+s}x\, h(x)\, \alpha^{(0)}_x
(W(f)) =: W_0
\end{equation}
in the strong-operator topology.
\end{Lemma}
{\it Proof of Lemma 3.2.}
(a) In view of the facts that $\mvsz \lcrc \sgl = \omega^{(\la m)}$
  and $\omega^{(0)}$ is invariant under the
action of the dilations $\sigma_{\la}$, statement (a) is equivalent
to 
\begin{equation}
\lim_{\la \to 0}\,||\,(\omega^{(m)} - \omega^{(0)})\rest
\Rr^{(0)}(\la \Oo)\,||
= 0\,.
\end{equation}
Since  one has $\bigcap_{\la > 0}
\Rr^{(0)}(\la\Oo)^- = \CC 1$
on general grounds \cite{Wigh}, relation (3.5) follows from
an argument by Roberts \cite{Rob1} because $\omega^{(m)}$ is locally
normal to $\omega^{(0)}$, cf.\ also \cite{BV}.
\\[4pt]
(b) By construction, $\Wu$ is obtained through convolution
(with respect to the lifted action
$\umalz_x$)
of the uniformly bounded function $\la \to \sgl(W(f))$
and a test-function $h$.  Thus it is strongly continuous with respect to
the action of $\umalz_x$. As a consequence of (2.13) and the support
 properties of $f$ and $h$  one observes that 
 $\Wu_{\la} \in \Rr^{(m)}(\la\Oo_0)$
for some double cone  $\Oo_0$ based on the time $t=0$ plane.
Hence, in view of the continuity properties of $\Wu$ 
with respect to the translations there holds   
$\Wu \in \uAAm(\Oo_0)$.

For the final part of the statement we note that for all $x =(t,\vx)
\in \RR^{1+s}$ one has
\begin{eqnarray}  & &
||\,(\tau^{(m)}_x -\tau^{(0)}_x)f\,||^2_0
\ = \ ||\,(\tau^{(m)}_t - \tau^{(0)}_t)f\,||^2_0 \ =  \\[4pt]
& & \hspace*{-7mm}
\int\!\! \left. \frac{d^sp}{2\mu_{0,\vp}}\right|
 (\cos(t\mu_{m,\vp}) - \cos(t \mu_{0,\vp}))\widetilde{{\rm Re}f}(\vp)
 -  ( \mu_{m,\vp}\sin(t\mu_{m,\vp})
- \mu_{0,\vp}\sin(t\mu_{0,\vp}))
\widetilde{{\rm Im}f}(\vp)  \nonumber \\
& & \hspace*{-7mm}+ \ i\mu_{0,\vp}(\cos(t\mu_{m,\vp}) -
      \cos(t\mu_{0,\vp}))\widetilde{{\rm Im}f}(\vp) 
 +  i\mu_{0,\vp}\!\! \left. \left(
    \frac{\sin(t\mu_{m,\vp})}{\mu_{m,\vp}}
    -  \frac{\sin(t \mu_{0,\vp})}{\mu_{0,\vp}} \right)
      \widetilde{{\rm Re}f}(\vp)\right|^2\,. \nonumber
\end{eqnarray}
By an application of the dominated convergence theorem one concludes
that one has 
$\lim_{\la \to 0}\,||\,(\tau_t^{(\la m)} -\tau^{(0)}_t)
f\,||^2_0
= 0$ uniformly in $t$ on compact intervals.
Employing a standard argument (e.g. \cite[Prop.\ 5.2.4]{BR}), this implies that
$W(\tau^{(\la m)}_x f)$
converges for $\la \to 0$, uniformly for $x$ in any compact subset of
$\RR^{1+s}$, in the strong operator topology to
$W(\tau^{(0)}_x f)$.
The claimed statement follows from that.
 \hfill $\Box$
\\[6pt]
Now we are in the position to show that the mapping $\phi$ given in (3.2)
is well-defined. Let $\Oo$ be any double cone.
Take $\Au \in \uAAm(\Oo)$ and an arbitrary $\Wu$ as in
(3.3).
Then consider
\begin{eqnarray}
\mvzi(\Wu\,\Au) & =& \lim_{\kappa}\, \omega^{(m)}(\Wu_{\lk}
\Au_{\lk}) \\
& =& \lim_{\kappa}\,\omega^{(\lk m)}(\sigma_{\lk}^{-1}(\Wu_{\lk})
\sigma_{\lk}^{-1}(\Au_{\lk})) \nonumber \\
& = & \lim_{\kappa}\,\omega^{(0)}(\sigma_{\lk}^{-1}(\Wu_{\lk})
\sigma_{\lk}^{-1}(\Au_{\lk})) \nonumber \\
& = & \lim_{\kappa}\,\omega^{(0)}(W_0\sigma_{\lk}^{-1}(\Au_{\lk}))\,.
\nonumber
\end{eqnarray}
Here, the second equality results from
$\omega^{(m)} \lcrc \sgl = \omega^{(\la m)}$, the third follows from
Lemma 3.2(a) and the fact that
$\sgl^{-1}(\Wu_{\la})\sgl^{-1}(\Aul) \in \Rr^{(0)}(\Oo_0)$ for
some fixed double cone $\Oo_0$ and all $\la > 0$,
 and the last equality is obtained from Lemma 3.2(b).
Now notice that the linear combinations of the $W_0$ of the form of (3.4),
 as $h$ and $f$
vary, are dense in $\Rr^{(0)}(\Oo)$. Since $\omega^{(0)}$ 
has the Reeh-Schlieder property (the corresponding GNS-vector is
cyclic and separating for all local algebras $\Rr^{(0)}(\Oo)$),
 it follows from (3.7) that
$w-\lim_{\kappa}\sigma_{\lk}^{-1}(\Au_{\lk})$ exists.
Moreover, since
\begin{equation} ||\,\poi(\Au)\Ooi\,||^2 = 
\lim_{\kappa}\,
\omega^{(0)}(\Au_{\lk}^*
\Au_{\lk}) =
\lim_{\kappa}\,
\omega^{(0)}(\sigma_{\lk}^{-1}(\Au_{\lk})^*\sigma^{-1}_{\lk}
(\Au_{\lk}))\,, 
\end{equation}
 the map $\phi$ in (3.2) maps 0 to 0 and hence is well-defined and linear.
 It is also *-preserving since the *-operation is continuous in the
 weak topology. One also has
\begin{equation}
 \langle \Ooi,\poi(\Au)\Ooi \rangle = \omega^{(0)}(\phi(\poi(\Au)))\,,
\quad \Au \in \uAAm\,,
\end{equation}
which already shows that $\phi$ connects $\mvoi$ and $\omega^{(0)}$.

In the next step, we will establish the intertwining relation 
\begin{equation}
\phi(\maoi_x(\poi(\Au))) = \alpha^{(0)}_x(\phi(\poi(\Au)))\,,\quad
\Au \in\uAAm\,.
\end{equation}
Let us pick arbitrarily $\Au \in \uAAm(\Oo)$, and
$\Wu$ as in (3.3).  Lemma 3.2(b) implies that
$ \lim_{\kappa}\,\alpha^{(\lk m)}_x \lcrc \sigma^{-1}_{\lk}(\Wu_{\lk}) =
\alpha^{(0)}_x(W_0)$ in the strong operator topology. Making use of
this we are led to the following chain of equations, valid for
each $x \in \RR^{1+s}$: 
\begin{eqnarray}
\omega^{(0)}(W_0 \sigma_{\lk}^{-1}(\alpha^{(m)}_{\lk x}(\Au_{\lk})))
 & = & \omega^{(0)}(W_0\alpha_x^{(\lk m)}(\sigma^{-1}_{\lk}(\Au_{\lk}))) \\
& = & \omega^{(0)}(\sigma_{\lk}^{-1}(\Wu_{\lk})
\alpha_x^{(\lk m)}(\sigma_{\lk}^{-1}(\Au_{\lk}))) + O(1)
\nonumber \\
& = &
\omega^{(\lk m)}(\sigma_{\lk}^{-1}(\Wu_{\lk})
\alpha_x^{(\lk m)}(\sigma_{\lk}^{-1}(\Au_{\lk}))) + O(1)
\nonumber \\
& = & \omega^{(\lk m)}(\alpha_{-x}^{(\lk m)} \lcrc
\sigma_{\lk}^{-1}(\Wu_{\lk})\sigma_{\lk}^{-1}(\Au_{\lk})) + O(1)
\nonumber \\
& = & \omega^{(0)}(\alpha_{-x}^{(\lk m)} \lcrc \sigma_{\lk}^{-1}
(\Wu_{\lk})\sigma_{\lk}^{-1}(\Au_{\lk})) + O(1) 
\nonumber \\
& = & \omega^{(0)}(\alpha^{(0)}_{-x}(W_0)\sigma^{-1}_{\lk}
(\Au_{\lk})) + O(1) \nonumber \\ &=&
\omega^{(0)}(W_0\alpha^{(0)}_x\lcrc \sigma_{\lk}^{-1}(\Au_{\lk})) + O(1)\,,
\nonumber
\end{eqnarray}
where in each line, $O(1)$ denotes some function of $\lk$ which tends
to $0$ as $\lk \to 0$. For the first equality in (3.11) we have used
eq.\ (2.8), the second
 follows from Lemma 3.2(b), the third from Lemma 3.2(a).
One passes to the fourth equality by invariance of the states
under the corresponding action of the dynamics, and to the
fifth by using again Lemma 3.2(a).
The sixth equation is obtained from Lemma 3.2(b), and the last equality
makes again use of the invariance of the state under the action
of the translations. The desired relation (3.10) then results in the
limit $\lk \to 0$, using again the fact that the span of the $W_0$ is dense
in $\Rr^{(0)}(\Oo)$, and that $\Omega^{(0)}$ has the Reeh-Schlieder
property.

In a similar way one can establish the corresponding intertwining
relations for the Lorentz transformations.

Now we want to show that the map $\phi$ is also multiplicative.
At this point we make essential use of the fact that the 
massless free scalar field theory in $1+s$\,-dimensional Minkowski
spacetime satisfies, for $s = 2,3$, the Haag-Swieca compactness
condition \cite{BJu}. We aim at proving 
\begin{Lemma}
 For all local  $\Au,\Bu \in \uAAm$ 
it holds that
 \begin{equation}
  \phi(\poi(\Au))\,\phi(\poi(\Bu)) = \phi(\poi(\Au)\poi(\Bu))\,.
\end{equation}
\end{Lemma}
{\it Proof of Lemma 3.3.} Suppose first that $\Au$,\,$\Bu$ are localized
in two separated double cones $\Oo_A$,$\Oo_B$ based on the time $t=0$ plane. 
Then we will show that
\begin{equation}
\omega^{(0)}(\phi(\poi(\Au))\,\phi(\poi(\Bu))) =
\omega^{(0)}(\phi(\poi(\Au)\poi(\Bu)))\,.
\end{equation}
To this end we make use of the following result in \cite{BDF} which is
a consequence of the positivity of energy: For any given
$\delta > 0$, there is some
continuous, rapidly decaying function $f : \RR \to \RR$ such that
\begin{equation}
\langle\Ome,AB\Ome \rangle = \langle \Ome, Af(H_0)B
\Ome \rangle + \langle \Ome,Bf(H_0)A\Ome\rangle
\end{equation}
holds for all pairs of operators $A,B$ which satisfy
$[\alpha_t^{(0)}(A),B] = 0$ for $-\delta < t < \delta$. Here,
$\Ome$ denotes the GNS-vector of $\omega^{(0)}$ 
and $H_0$ the generator of the zero-mass dynamics (time-translations).
 Now $\Ak := \sigma_{\lk}^{-1}(\Au_{\lk}) \in \Rr^{(0)}(\Oo_A)$
and $\Bk := \sigma_{\lk}^{-1}(\Bk) \in \Rr^{(0)}(\Oo_B)$ and
consequently, because of locality, there is some $\delta > 0$ such
that $[\alpha^{(0)}_t(\Ak),B_{\kappa'}] = 0$ for $|t| < \delta$ and
all $\kappa,\kappa'$. 
 Thus
the aforementioned result applies to the effect that 
\begin{eqnarray}  & &
\omega^{(0)}(\phi(\poi(\Au))\phi(\poi(\Bu)))  = 
\lim_{\kappa}\,\lim_{\kappa'}\,\omega^{(0)}(\Ak B_{\kappa'}) \\
& & = \lim_{\kappa}\,\lim_{\kappa'}\,
\left( \langle\Ome,\Ak f(H_0)B_{\kappa'}\Ome\rangle
+ \langle \Ome,B_{\kappa'} f(H_0)\Ak\Ome\rangle \right)
\nonumber
\end{eqnarray}
for some $f$ as above. The crucial point
here is that, since $w-\lim_{\kappa'}B_{\kappa'}$ exists
and since $B_{\kappa'} \in \Rr^{(0)}(\Oo_B)$, there exist also the
strong limits $s-\lim_{\kappa'}f(H_0)B_{\kappa'}\Ome$ and
$s-\lim_{\kappa'}f(H_0)^*B_{\kappa'}^*\Ome$ because of the 
Haag-Swieca compactness for the massless free scalar field. This entails
that one may pass from the double limit in (3.15) to the associated diagonal
limit, i.e.\ 
\begin{equation}
\lim_{\kappa}\,\left(
\langle\Ome,\Ak f(H_0)\Bk\Ome\rangle +
\langle\Ome,\Bk f(H_0)\Ak\Ome\rangle \right)
\end{equation}
exists and equals the expression on the right hand side of (3.15).
Since
\begin{equation}
\langle\Ome,\Ak f(H_0)\Bk\Ome\rangle +
\langle\Ome,\Bk f(H_0)\Ak\Ome\rangle = \omega^{(0)}(\Ak\Bk) =
\omega^{(0)}(\sigma_{\lk}(\Au_{\lk}\Bu_{\lk}))\,, 
\end{equation}
relation (3.13) is thus established.

In a second step, we shall extend the relation (3.13) to
arbitrary local operators $\Au,\Bu$. Here we make use of the general
fact that any scaling limit theory is translation covariant and 
fulfills the spectrum condition \cite{BV}. For the case at hand this
implies that the function
\begin{equation}
\RR^{1+s} \owns x \mapsto G(x) :=
\ooi(\Au\,\umalz_x(\Bu))
\end{equation}
is the boundary value of an analytic function in the forward
tube $\RR^{1+s} + iV_+$. The same holds for the function
\begin{equation}
\RR^{1+s} \owns x \mapsto F(x) := \langle \Ome,
\phi(\poi(\Au))\alpha_x^{(0)}(\phi(\poi(\Bu)))\Ome\rangle\,. 
\end{equation}
Since the operators $\Au$ and $\umalz_x(\Bu)$ are, for sufficiently
large spacelike $x$, localized in disjoint double cones 
$\Oo_A$ and $\Oo_B$ as in the preceding step,
we may then appeal to (3.13) to conclude that
\begin{eqnarray}
F(x) & = & \langle \Ome,\phi(\poi(\Au))\phi(\poi(\umalz_{x}(\Bu)))
\Ome\rangle \\
& = & \langle \Ome,\phi(\poi(\Au)\poi(\umalz_{x}(\Bu)))\Ome
\rangle \nonumber \\
& = & \langle\Ooi,\poi(\Au\,\umalz_{x}(\Bu))\Ooi \rangle \ = \
G(x) \nonumber
\end{eqnarray}
for some open set of translations $x$. By analyticity, we then
obtain $F(x) = G(x)$ for all $x \in \RR^s$, implying
(3.13) for arbitrary  local operators $\Au,\Bu$.

Finally, (3.13) must be generalized to the operator identity (3.12).
Let $\Wu$ be of the form (3.3).
 Due to the strong-operator convergence of the
families $\sigma_{\la}^{-1}(\Wu_{\la})$ as $\la \to 0$, there holds
for each $\Au \in \uAAm(\Oo)$ the following 
restricted form of multiplicativity
of the map $\phi$:
\begin{equation}
 \phi(\poi(\Wu))\phi(\poi(\Au)) = \phi(\poi(\Wu)\poi(\Au)) 
\end{equation}
and similarly
\begin{equation}
 \phi(\poi(\Au))\phi(\poi(\Wu)) = \phi(\poi(\Au)\poi(\Wu)) \,.
\end{equation}
Therefore, when $\Wu'$ is also of the form (3.3) and when
the $\Au$ and $\Bu$ in (3.13) are replaced by $\Wu\cdot\Au$
and $\Bu\cdot\Wu'$, respectively,
we conclude that with $W_0,W_0'$ as in (3.4)
\begin{eqnarray}
\omega^{(0)}(W_0\phi(\poi(\Au))\phi(\poi(\Bu))W_0') =
\omega^{(0)}(\phi(\poi(\Wu\cdot\Au))\phi(\poi(\Bu\cdot\Wu')))\\
 = \ \omega^{(0)}(\phi(\poi(\Wu\cdot\Au))\poi(\Bu\cdot\Wu')) = 
\omega^{(0)}(W_0\phi(\poi(\Au)\poi(\Bu))W_0')\,. \nonumber
\end{eqnarray}
 Then the equation (3.12)
results again from  the fact that the linear
span of the elements $W_0,W_0'$  is dense in the local von Neumann algebras
$\Rr^{(0)}(\Oo)$, and from the Reeh-Schlieder property of $\omega^{(0)}$.
 This completes the proof of
the multiplicativity of the map $\phi$.
\hfill  $\Box$ 

It remains to be shown that $\phi$ maps $\mAoi(\Oo)$ onto $\Aa^{(0)}(\Oo)$
for all double cones $\Oo$. We will prove this first 
for all $\Oo_0$ based on the time $t=0$ plane. 

Let $A \in \Aa^{(0)}(\Oo_0)$,
 choose some function $h \in {\cal D}(\RR^{1+s})$, and
define the function $\RR^+ \owns \la \mapsto \Au^{(h)}_{\la}$
by
\begin{equation}
\Au^{(h)}_{\la} := \int d^{1+s}x\,h(x)\,\alpha^{(m)}_{\la x}(\sigma_{\la}
(A))\,. 
\end{equation}
Then $\Au^{(h)} \in \uAAm(\Oo_h)$ where $\Oo_h$
 shrinks to $\Oo_0$ when
the support of $h$ is shrunk  to $\{0\}$.
We observe that
\begin{equation}
\sigma_{\la}^{-1}(\Au^{(h)}_{\la}) = \int d^{1+s}x\,h(x)\,
\alpha^{(\la m)}_x(A)\,, 
\end{equation}
from which we now obtain the following chain of equations, where
$\Wu \in \uAAm$ is of the form (3.3), and $W_0$ relates to it
as in (3.4).
\begin{eqnarray}
\lim_{\la \to 0}\,\omega^{(0)}(W_0\sigma_{\la}^{-1}(\Au^{(h)}_{\la}))
& = & \lim_{\la \to 0}\,\omega^{(0)}(\sigma_{\la}^{-1}(\Wu_{\la})
\sigma_{\la}^{-1}(\Au^{(h)}_{\la})) \\
& = & \lim_{\la \to 0}\,\omega^{(\la m)}(\sigma^{-1}_{\la}(\Wu_{\la})
\sigma_{\la}^{-1}(\Au^{(h)}_{\la})) \nonumber\\
& = & \lim_{\la \to 0}\, \int d^{1+s}x\,h(x)\,\omega^{(\la m)}
(\alpha^{(\la m)}_{-x}\lcrc \sigma_{\la}^{-1}(\Wu_{\la})\cdot A)\nonumber
\\
& = & \lim_{\la \to 0}\, \int d^{1+s}x\,h(x)\,\omega^{(0)}
(\alpha^{(\la m)}_{-x}\lcrc \sigma_{\la}^{-1}(\Wu_{\la})\cdot A)\nonumber\\
& = & \int d^{1+s}x\,h(x)\,\omega^{(0)}(\alpha_{-x}^{(0)}(W_0)A)\,.\nonumber
\end{eqnarray}
In the preceding chain, the first equality is derived by means of
Lemma 3.2(b), the second by
Lemma 3.2(a), and the third uses the invariance of
$\omega^{(\la m)}$ under the translations $\alpha^{(\la m)}_x$. The
fourth equation is implied by 
Lemma 3.2(a)
 and the last one follows on
account of Lemma 3.2(b). Falling back on the by now familiar argument
that the span of elements $W_0$ 
 is weakly dense in $\Rr^{(0)}(\Oo_0)$, the just obtained
equalities entail that 
\begin{equation}
\phi(\poi(\Au^{(h)})) = w-\lim_{\la \to 0}\,\sigma_{\la}^{-1}(\Au^{(h)}
_{\la}) = \int d^{1+s}x\, h(x)\,\alpha_x^{(0)}(A)\,. 
\end{equation}
Since $A$ and $h$ were arbitrary
we see that $\phi(\mAoi)\Omega$ spans $\Hh^{(0)}$, the standard
Hilbert space. Hence, because of the algebraic properties
  of $\phi$ and
(3.9), $\phi$ is in fact given by the adjoint action of a
unitary $\Hoi \to \Hh^{(0)}$ defined through the assignment
$\poi(\Au)\Ooi \mapsto \phi(\poi(\Au))\Omega$.
Therefore, $\phi$ is in particular injective.

 If one takes now
a sequence $h_n$ of test-functions approaching the Dirac-measure
$\delta$, one has
$\phi(\poi(\Au^{(h_n)}))
 \to A$ for $n \to \infty$ in norm since
$A \in \Aa^{(0)}(\Oo_0)$ is, by the very definition of the algebra
$\Aa^{(0)}(\Oo_0)$, an element on which the translations
$\alpha^{(0)}_x$, $x \in \RR^{1+s}$, act strongly continuously.
This shows that, for each pair $\Oo_0,\Oo_1$ of
double cones based on the time $t=0$ hyperplane
 with $\overline{\Oo_0} \subset \Oo_1$, we have
$\Aa^{(0)}(\Oo_0) \subset \phi(\mAoi(\Oo_1))$. On the other hand,
for each $\Oo_1$ there holds the inclusion
$\phi(\mAoi(\Oo_1)) \subset \Aa^{(0)}(\Oo_1)$  in view of the
intertwining relation (3.10) and the fact that $\phi(\mAoi(\Oo_1))
\subset \Rr^{(0)}(\Oo_1)$. Hence
\begin{equation}
\Aa^{(0)}(\Oo_0) \subset \phi(\mAoi(\Oo_1)) \subset
\Aa^{(0)}(\Oo_1)\,, 
\end{equation}
and because of the injectivity of $\phi$ we conclude that
\begin{equation}
\Aa^{(0)}(\Oo_0) \subset \phi \left(\bigcap_{\Oo_1 \supset 
\overline{\Oo_0}}
\mAoi(\Oo_1)\right) \subset  \bigcap_{\Oo_1 \supset \overline{\Oo_0}}
\Aa^{(0)}(\Oo_1)\,. 
\end{equation}
The required equality $\phi(\mAoi(\Oo_0)) = \Aa^{(0)}(\Oo_0)$ 
 follows now from  the outer continuity of the nets
$\Oo \to \Aa^{(0)}(\Oo)$ and $\Oo \to \mAoi(\Oo)$.
Due to the intertwining property  of $\phi$ for the Poincar\'e 
transformations, cf.\ (3.10), we immediately obtain
\begin{eqnarray}
\phi(\mAoi(\Lambda\Oo_0 + x)) =  \phi(\maoi_{\Lx}(\mAoi(\Oo_0))) \\
 = \   \alpha^{(0)}_{\Lx}(\phi(\mAoi(\Oo_0))) 
 =  \Aa^{(0)}(\Lambda\Oo_0 + x) \nonumber
\end{eqnarray}
for all $\Oo_0$ based on the time $t=0$ plane and all $\Lx$.
 Consequently the equality
  $\phi(\mAoi(\Oo)) = \Aa^{(0)}(\Oo)$ holds also for
all double cones $\Oo$.
\\[6pt]
This completes the proof of Theorem 3.1. 
%%%%%%%%%%%%%%%%%%%%%%%%%%%%%%%%%%%%%%%%%%%%%%%%%%%%%%%%%%%%%%%%%%%%%%%%%
\section{Construction of charged states for $s = 1$}
\setcounter{equation}{0}
%% \setcounter{Theorem}{0}
%%%%%%%%%%%%%%%%%%%%%%%%%%%%%%%%%%%%%%%%%%%%%%%%%%%%%%%%%%%%%%%%%%%%%%%%%
As we shall see, the properties of the scaling limit theory of the
free massive scalar field  in two-dimensional
Minkowski-spacetime  are in several respects different from  the case in 
three- and four-dimensional Minkowski-spacetime.
First, while the vacuum states appearing in the scaling limit are
always pure in $s \ge2$ dimensions \cite{BV}, the present model in
$s=1$ dimension provides an example where these states are mixed.
Second, there appear charged states in the scaling limit whose
restrictions to the algebras of both, the right and left component of
the spacelike complement of a double cone region, coincide with the
vacuum.
But their restriction to the algebra of the full spacelike complement
is disjoint from the vacuum.
Hence these states carry a gauge charge in the sense of
\cite{BF}. As was discussed in \cite{Bu96}, there holds Gauss' law for
this charge. These results will be formulated more precisely in the
following theorem.
\begin{Theorem}
Let $\mvzi$ be any scaling limit state of the theory of the free scalar
field of mass $m > 0$ in two-dimensional Minkowski-spacetime. Then for
the corresponding scaling limit theory
$(\mAoi,\mazi,\mvzi)$ it holds that:
\\[6pt]
(a) \quad $\mAoi$ possesses a non-trivial center.
\\[6pt]
(b) \quad There are  charged
states $\oip$ on $\mAoi$ which are locally normal
to $\mvzi$ and have moreover the following
properties:\footnote{As usual, the
  $C^*$-algebras corresponding to unbounded regions are generated by
  the algebras associated with all double cones which are contained in
  the respective region.}\\[6pt]
${}$\,\,(b.i)\quad $\oip \rest \mAoi(\Oo^{(\pm)})= \mvzi \rest
 \mAoi(\Oo^{(\pm)})$ 
for  sufficiently large double cones $\Oo$, where $\Oo^{(\pm)}$
denotes the right/left component of $\Oo'$.
\\[4pt]
${}$\,(b.ii)\quad $\oip \rest \mAoi(\Oo')$
is disjoint from $\mvzi \rest \mAoi(\Oo')$ for all double cones
$\Oo$.\\[4pt]
(b.iii)\quad In the GNS-representation induced by $\oip$ the
translations $\mazi_{\RRf^2}$ are implemented by a continuous unitary
representation satisfying the spectrum condition.
\end{Theorem}
The remainder of this section is devoted to the proof of these
statements.\vspace{3mm}

Let $m > 0$ be given. We choose the Fock-representation induced by
$\omega^{(m)}$ as our standard representation. The corresponding
GNS-vacuum vector will be denoted by $\Omega^{(m)}$.

To prove (a), we define for arbitrary $h \in {\cal D}(\RR^2)$ and
real-valued $f \in \DR$,
\begin{equation}
 \Cu^{(h)}_{\la} := \int d^2x\, h(x)\,\alpha^{(m)}_{\la x}\left( W(|\ln
 \la|^{-1/2}\dl f)\right)\,, \quad \la > 0\,.
\end{equation}
Then $\la \mapsto \Cu^{(h)}_{\la}$ is contained in $\uAAm(\Oo)$ for a
suitable double cone $\Oo$. Moreover, there hold the following
relations:
\begin{eqnarray}
\lim_{\la \to 0} \,\langle \Omega^{(m)},\Cu^{(h)}_{\la}
\Omega^{(m)}\rangle
& = &  \int d^2x \,h(x) \cdot {\rm e}^{-(1/2)
  |\tilde{f}(0)|^2}
\,,\\  
\lim_{\la \to 0} \,\langle \Omega^{(m)},\Cu^{(h)}_{\la}\Cu^{(h)}_{\la}
\Omega^{(m)}\rangle
& = & \left( \int d^2x \,h(x)\right)^2 \cdot {\rm e}^{-2\,
  |\tilde{f}(0)|^2}
\,,\\   
\lim_{\la \to 0} \,\langle \Omega^{(m)},\Cu^{(h)}_{\la}{}^*\Cu^{(h)}_{\la}
\Omega^{(m)}\rangle
 & = & \left|\, \int d^2x \,h(x)\,\right|^2\,.
\end{eqnarray}
Equation (4.2) can be derived as follows. Inserting the expression  
for $\Cu^{(h)}$ into the vacuum state yields
\begin{equation}
\langle \Omega^{(m)},\Cu^{(h)}_{\la}
\Omega^{(m)}\rangle = \int d^2x \,h(x)\cdot
{\rm e}^{-(1/2)||\,|\ln \la |^{-1/2}\dl f\,||_m^2}\,,
\end{equation}
where we used the invariance of the vacuum under translations. 
 For the  term in the exponential we obtain by partial integration 
\begin{eqnarray}
|| \, | \ln \la |^{-1/2}\dl f\,||_m^2  
= \ |\ln \la |^{-1}\cdot \int_{0}^{\infty}\frac{d\vp}
{\sqrt{\vp^2 + (\la m)^2}}\,|\tilde{f}(\vp)|^2  \hspace*{20mm} \\
= - |\ln \la |^{-1} \cdot  \ln (\la m) \,|\tilde{f}(0)|^2 - 
  |\ln \la|^{-1}\cdot\int_0^{\infty}d\vp
 \ln \big(
 \vp + \sqrt{\vp^2 + (\la
 m)^2} \big) \,\frac{d}{d\vp}|\tilde{f}(\vp)|^2  \,, \nonumber
\end{eqnarray}
where we made use of the fact that $\vp \mapsto |\tilde{f}(\vp)|^2$ is
symmetric since $f$ is real. Since $f$ is a test-function, relation
(4.2) then follows. For the proof of (4.3) and (4.4) one has also to make
use of the Weyl-relations and the specific form of the action of the
translations on the Weyl-operators. Otherwise the reasoning is
completely analogous.

Now let $(\poi,\Hoi,\Ooi)$ denote the GNS-representation of
$\mvzi$, and $\Coi^{(h)} := \poi(\Cu^{(h)})$.
We want to show that $\Coi^{(h)}$ is contained in the center of
$\mAoi$ (for any choice of $h$ and $f$). Due to locality it is
sufficient to show that $\Coi^{(h)}$ is translation invariant, i.e.\
$\maoi_{x}(\Coi^{(h)}) = \Coi^{(h)}$ for all $x$. From the definition
of $\Cu^{(h)}$ it follows that $\umalf_x(\Cu^{(h)}) = \Cu^{(h_x)}$
where $h_x(y) = h(y -x)$ denotes the translate of $h$. Moreover, it
is easily seen that $\Cu^{(h_x)} - \Cu^{(h)} = \Cu^{(h_x - h)}$.
 Hence we have, by (4.4),
\begin{eqnarray}
||\, (\maoi_x(\Coi^{(h)}) - \Coi^{(h)})\Ooi\,||^2 
& = & ||\,\Coi^{(h_x - h)}\Ooi\,||^2 \\
& = & \lim_{\la \to 0}\langle \Omega^{(m)},\Cu^{(h_x-h)}_{\la}{}^*
\Cu^{(h_x-h)}_{\la}\Omega^{(m)} \rangle \nonumber \\
& = & \left|\,\int d^2y\,\left( h_x(y) - h(y)\right)\,\right|^2 \ = \ 
0\,. \nonumber
\end{eqnarray}
Since $\Coi^{(h_x-h)}$ is contained in $\mAoi(\Oo_x)$ for some double
cone $\Oo_x$, and since $\Ooi$ is separating for these local algebras
\cite[Lemma 6.1]{BV}, the last equality entails that $\Coi^{(h_x -h)}
= 0$. Thus
$\Coi^{(h)}$ is invariant under translations and hence lies in the
center of $\mAoi$.
Finally, (4.2) and (4.3) imply that $\Coi^{(h)}$ is different from a
multiple of 1 in the case that $\int d^2x\,h(x) \neq 0$ and
$\int d\vx\, f(\vx) \neq 0$, because then one has
\begin{eqnarray}
& &\langle \Ooi,\Coi^{(h)}\Ooi\rangle^2
\ =\ \left(\, \int d^2x\,h(x)\,\right)^2\cdot {\rm
 e}^{-|\tilde{f}(0)|^2}\\
& &  \quad \neq\ \left(\, \int d^2x\,h(x)\,\right)^2\cdot {\rm
 e}^{-2\,|\tilde{f}(0)|^2}
\ =\ 
\langle\Ooi,\Coi^{(h)}\Coi^{(h)}\Ooi \rangle\,. \nonumber
\end{eqnarray}
Part (a) of the theorem is thereby proved. 

 We note that the appearance of a non-trivial center in
the scaling limit may be attributed to the fact that the massive free field
in two dimensions contains an anomalous classical part, cf.\ 
\cite{Bu96}.

Let us now turn to the proof of part (b) of the theorem.
 For given $h \in {\cal D}(\RR^2)$
and $f \in \DR$ we define $\Wu_{\la}(f) := \sgl(W(f))$ and
\begin{equation}
 \Wu^{(h)}_{\la}(f)  :=\int d^2x\,h(x)\,\alpha^{(m)}_{\la
 x}(\Wu_{\la}(f)) \, . 
\end{equation}
Then  $\la \mapsto \Wu^{(h)}_{\la}(f)$
is contained in $\uAAm(\Oo)$ for some double cone $\Oo$. 
Whereas the scaling algebra contains only these regularized
 Weyl-operators, one can recover from them the non-regularized
 Weyl-operators in the scaling limit. This is shown in the subsequent lemma.
\begin{Lemma} Let $(\poi,\Hoi,\Ooi)$
 be the GNS-representation of $\mvzi$, and  let $h_n$ be any sequence
  of smooth functions on $\RR^2$ with
$h_n(x) \ge 0$, $h_n(x) = 0$ for $|x| \ge n^{-1}$ and $\int d^2x\,h_n(x) = 1$.
 Then for each $f \in \DR$,
\begin{equation}
 \Woi(f) := \lim_{n \to \infty}\,\poi
(\Wu^{(h_n)}(f))
\end{equation}
exists in the strong operator topology and, for the net $\lk$, $\kappa
\in \KK$, used in the construction of $\mvzi$,
\begin{equation}
 \lim_{\kappa}\,\omega^{(m)}(\Wu_{\lk}(f)\Au_{\lk}\Wu_{\lk}
(g)) = \langle \Ooi, \Woi(f)\poi(\Au)
 \Woi(g)\Ooi \rangle
\end{equation}
for all $f,g \in \DR$, $\Au \in \uAAm$. Moreover, there hold the 
Weyl relations
\begin{equation}
\Woi(f)\Woi(g)
= {\rm e}^{-i\,\sigma(f,g)/2}
\Woi(f+g)\,, \quad
f,g \in \DR\,.
\end{equation}
\end{Lemma}
{\it Proof of Lemma 4.2}.
First we show that for $f \in \DR$ one has
\begin{equation}
\sup_{1 > \la > 0}\,||\,(\alpha^{(m)}_{\la x}
 (\Wu_{\la}(f)) - \Wu_{\la}(f))\Omega^{(m)}\,||
\to 0 \quad {\rm for}\quad  x \to 0 \,.
\end{equation}
Since $\omega^{(m)} \lcrc \sgl = \omega^{(\la m)}$ and in view of
(2.8),
this amounts to showing that
\begin{equation}
\sup_{1 > \la > 0}\,||\,(\alpha^{(\la m)}_{x}
(W(f)) - W(f))\Omega^{(\la m)}\,||
\to 0 \quad {\rm for}\quad  x \to 0 \,.
\end{equation}
For the latter it is sufficient to show that
\begin{equation}
\sup_{1 > \la > 0}\,||\, \tau_x^{(\la m)}  f - f \,||_{\la m}
\to 0 \quad {\rm for} \quad x \to 0 \,.
\end{equation}
But for all $x = (t,\vx)\in \RR^2$ and $1 > \la > 0$ one obtains 
\begin{eqnarray}
& & ||\, 
 \tau^{(\la m)}_x f - f \,||^2_{\la m} \\
& & =\ 
\int \frac{d\vp}{2\mu_{\la m,\vp}}
 \left| ({\rm e}^{it\mu_{\la m,\vp} -i\vp\vx}-1)
\left(\widetilde{{\rm Re}f}(\vp) + i\mu_{\la m,\vp}
\widetilde{{\rm Im}f}(\vp)\right)\right|^2  \nonumber \\
& & \le \
2 \int d\vp\, (|t| + |x|)^2\mu_{m,\vp}
\left( |\widetilde{{\rm Re}f}(\vp)| 
 +  \mu_{ m,\vp}|\widetilde{{\rm Im}f}(\vp)| \right)^2
 \
 \le \
 {\rm const}\, (|t| + |x|)^2,  \nonumber
\end{eqnarray}
hence (4.13) is established.

Now let $f \in \DR$ be given. For $\epsilon > 0$,
we choose some $n \in \NN$ so that
\begin{equation}
\sup_{1 > \la > 0}\,||\,(\alpha^{(m)}_{\la x}(
 \Wu_{\la}(f)) - \Wu_{\la}(f))\Omega^{(m)}\,|| < 
\epsilon/4
\end{equation}
for  $|x| < n^{-1}$. Then for all $j,k > n$ we have
\begin{eqnarray}
& & ||\, (\poi(\Wu^{(h_j)}_{\la}(f)) - \Wu^{(h_k)}_{\la}(f))
\Ooi \,|| \\
& & = \
\lim_{\kappa}\,||\,(\Wu^{(h_j)}_{\lk}(f) - \Wu^{(h_k)}_{\lk}(f))
\Omega^{(m)} \,|| 
 \
\le \   \sup_{1 > \la > 0}\,||\,(\Wu^{(h_j)}_{\la}(f) -
 \Wu^{(h_k)}_{\la}(f))
\Omega^{(m)} \,|| \nonumber \\
& & \
\le 2 \int d^2x \,(h_j(x) +h_k(x))  
 \sup_{1 > \la > 0}\,||\,(\alpha^{(m)}_{\la x}(
 \Wu_{\la}(f)) - \Wu_{\la}(f))\Omega^{(m)}\,||
< \epsilon\,. \nonumber
\end{eqnarray}
Here we used that the integral equals 2 and the support of $h_j$ is
contained in the ball of radius $n^{-1}$ for $j > n$.
Since $\Ooi$ is separating for the local von Neumann algebras 
$\mAoi(\Oo)^-$, this last estimate shows that $\poi(\Wu^{(h_j)}(f))$
is strongly convergent to an element in $\mAoi(\Oo)^-$ for some double
cone $\Oo$ as $j \to \infty$.

To prove (4.11), we note that (4.17) also implies that
$\sup_{1 > \la > 0}\,||\,
(\Wu^{(h_j)}_{\la}(f) -
 \Wu_{\la}(f))
\Omega \,|| < \epsilon/2$,
  for all
$j > n$. Hence one obtains that
\begin{equation}
\sup_{1 > \la > 0}\, |\, 
 \omega^{(m)}(\Wu^{(h_j)}_{\la}(f)\Aul\Wu^{(h_j)}_{\la}(g))
-
\omega^{(m)}(\Wu_{\la}(f)\Aul\Wu_{\la}(g))\,|
\end{equation}
can be made arbitrarily small for sufficiently large $j$. Relation
(4.11) is implied by this fact.

Now let us turn to proving (4.12). Let $f,g \in \DR$,
 then $\Woi(f)$ and $\Woi(g)$
are contained in $\mAoi(\Oo)^-$ for some double cone $\Oo$. With
the help of (4.11), when $\Oo_1 \in \Oo'$, we obtain
for all $\Au \in \uAAm(\Oo_1)$,
\begin{eqnarray}
 \langle \Ooi,\poi(\Au)\Woi(f)\Woi(g)\Ooi
 \rangle 
 &=& \langle \Ooi,\Woi(f)\poi(\Au)\Woi(g)\Ooi
 \rangle \\
&  = &
\lim_{\kappa}\,
\omega^{(m)}(\Wu_{\lk}(f)\Au_{\lk}\Wu_{\lk}(g))  \nonumber \\
& =& 
\lim_{\kappa}\,
\omega^{(m)}(\Au_{\lk}\Wu_{\lk}(f)\Wu_{\lk}(g)) \nonumber  \\
&= & 
\lim_{\kappa}\,
\omega^{(m)}(\Au_{\lk}{\rm e}^{-i\,\sigma(f,g)/2}
\Wu_{\lk}(f+g)) \nonumber  \\  
&=&  \langle \Ooi,\poi(\Au)
{\rm e}^{-i\,\sigma(f,g)/2}
\Woi(f+g)\Ooi
 \rangle \,, \nonumber
\end{eqnarray}
where we made use of locality and the Weyl-relations.
Since $\Ooi$ is cyclic for the $C^*$-algebra generated by
$\{\mAoi(\Oo_1),\, \Oo_1 \subset \Oo'\}$ \cite[Lemma 6.1]{BV}
and separating for $\mAoi(\Oo)^-$ 
this establishes (4.12).\hfill  $\Box$
\\[10pt]
On the basis of the last lemma, we can define the following two
families of unitary operators on the scaling-limit Hilbert space
$\Hoi$.
First, we pick some sequence of real-valued $f_n \in \DR$ where
$f_n(\vx) =q$, 
 $|\vx| < n$, for an arbitrary real number  $q$ and define
\begin{equation}
 \Yqn := \Woi(if_n)\,, \quad n \in \NN\,. 
\end{equation}
Secondly, for any choice of a real-valued $w \in \DR$ with $w(\vx) = 1$
for $|\vx| \le 1$, we consider the sequence of test-functions
$g_n(\vx) := \partial_{\vx}w(\vx/n)$, $n \in \NN$, and define 
\begin{equation}
 Z^{(n)}_w(r) := \Woi(r \, g_n)\,, \quad r \in \RR\,. 
\end{equation}
In the following lemma we collect
some properties of these unitaries.
\begin{Lemma}
It holds that
\begin{equation}
\lim_{n \to \infty}\, \langle \Yqn\Ooi,\poi(\Au)\Yqn\Ooi \rangle =
\langle \Ooi,\poi(\Au)\Ooi \rangle\,, \quad \Au  \in \uAAm\,,
\end{equation}
and
\begin{equation}
\langle\Ooi,Z^{(n)}_w(r)\Ooi\rangle = {\rm e}^{-(r^2/4) \cdot \int
d\vk\,|\vk|\,|\tilde{w}(\vk)|^2}\,, \quad r\in \RR\,,\ \  n \in \NN\,.
\end{equation}
\end{Lemma}
{\it Proof of Lemma 4.3.}
Let $a > 0$ and
$\Au \in \uAAm(\Oo_a)$, where $\Oo_a$ is the double cone based on the
interval $(-a,a)$ in the time $t=0$ plane. For $n > a$ we obtain
by (4.11) 
\begin{eqnarray}
\langle \Yqn\Ooi,\poi(\Au)\Yqn\Ooi \rangle & = &
\lim_{\kappa}\,\omega^{(m)}(\Wu_{\lk}(if_n)^*\Au_{\lk}\Wu_{\lk}
(if_n)) \\ & = &
\lim_{\kappa}\,\omega^{(m)}(W(if_n)^*\Au_{\lk}W(if_n))
\nonumber \\
& = & \langle \Ooi,\poi(\Au)\Ooi \rangle\,, \nonumber
\end{eqnarray}
where in the second equality we have used that
\begin{eqnarray}
\Wu_{\la}(if_n)^*\Aul\Wu_{\la}(if_n) &=&
W(\dl i f_n)^*\Aul W(\dl i f_n) \\  & = &
W(if_n)^*\Aul W(if_n)
\nonumber 
\end{eqnarray}
for $1 > \la >0$. This holds because both $if_n$
and $\dl if_n$ are equal to $iq$ in a neighborhood of the closure of
$(-\la a,\la a)$, and so we see that $if_n = i\chi_n^{(\la)} + \dl i f_n$
with a real-valued function $\chi_n^{(\la)} \in \DR$ having
support outside of $(-\la a,\la a)$. As $\Aul \in
\uAAm(\la \Oo_a)$, it commutes with
$W(i\chi_n^{(\la)})$ and thus the last identity of (4.26)
results from the Weyl-relations. This argument may also be used to
show that
$W(if_n)^*\Aul W(if_n)$, $1> \la >0$, is independent
of $n > a$. The last equality of (4.25) is then obtained  for
fixed $n > a$ since the scaling limit of a state locally normal to
$\omega^{(m)}$ is equal to the scaling limit of $\omega^{(m)}$,
 cf.\ \cite[Cor.\ 4.2]{BV}. This proves (4.23).

Equation (4.24) is obtained with the help of (4.11) by a straightforward
computation,
\begin{eqnarray} {}\quad \quad \quad \quad \quad \quad \ \ 
\langle \Ooi,Z^{(n)}_w(r)\Ooi \rangle & = &
\lim_{\kappa}\,\omega^{(m)}(\Wu_{\lk}(g_n)) \\
& = & \lim_{\kappa}\,{\rm e}^{-\frac{1}{2}
 ||\,g_n
\,||^2_{\lk m}} \nonumber \\
& = & {\rm e}^{-(r^2/4)\cdot \int d\vk\,|\vk|\,|\tilde{w}(\vk)|^2}\,.
\hspace*{44mm} \Box \nonumber
\end{eqnarray} 
We note that by construction $g_n(\vx) = 0$ for $|\vx| < n$, hence
 each weak limit point of $Z^{(n)}(r)$, as $n \to \infty$,
lies in the center of $\mAoi{}^-$.
Drawing on (4.24) it is straightforward to show that these central
elements are also different from multiples of the identity.

 We construct now a state $\oip$
with the properties as claimed in part (b) of the Theorem. For this
purpose, we
consider the operators
\begin{equation}
\Vqn := \Woi(iu_n)\, , \quad n \in \NN\,,
\end{equation}
where the sequence of real-valued $u_n \in \DR$ is required to have,
for some fixed $a > 0$, the following properties:
\begin{equation}
 u_n(\vx) =
\begin{cases}
 {}\quad\ \ \quad 0\,, & \vx \le -a\,,\\
 \text{independent of $n$,} & |\vx| \le a\,,\\
 { }\quad\ \ \quad q\,, & a \le \vx \le n\,a\,. 
\end{cases} 
\end{equation}
Here $q \neq 0$ is some real number (``charge'') which will be kept
fixed in the following.
With the help of the operators $\Vqn$ we define the state
\begin{equation}
\oip(\poi(\Au)) := \lim_n\, \langle
\Vqn\Ooi,\poi(\Au)\Vqn \Ooi
\rangle\,, \quad \Au \in \uAAm\,.
\end{equation}
 An argument similar to the one used to prove (4.23) shows that for each
double cone $\Oo$ in $\RR^2$ there exists some number $n_0$ such that,
for each $\Au \in \uAAm(\Oo)$, the expression
$\langle \Vqn\Ooi,\poi(\Au)\Vqn\Ooi
\rangle$ is independent of $n$ for $n > n_0$. Therefore, the state
$\oip$
exists and is  locally normal to the scaling limit vacuum
$\mvzi$.

 If $\Oo$ is a double cone which is contained in $\Oo_a^{(+)}$,
we obtain for all $\Au \in \uAAm(\Oo)$ and  sufficiently large $n$,
\begin{equation}
 \langle \Vqn\Ooi,\poi(\Au)\Vqn\Ooi
\rangle = \langle \Yqn\Ooi,\poi(\Au)\Yqn\Ooi
\rangle\,. 
\end{equation}
Again, this can easily be checked by an argument similar to that establishing
(4.23). In view of Lemma 4.3 we conclude that
\begin{equation}
 \oip \rest \mAoi(\Oo_a^{(+)})= \mvzi \rest
 \mAoi(\Oo_a^{(+)})\,. 
\end{equation}
On the other hand, if $\Oo$ is contained in 
 $\Oo_a^{(-)}$ it follows from locality that 
\begin{equation}
 \langle \Vqn\Ooi,\poi(\Au)\Vqn\Ooi
\rangle = \langle \Ooi,\poi(\Au)\Ooi
\rangle\,. 
\end{equation}
Hence relation (4.32) holds also with $\Oo^{(+)}$ replaced by $\Oo^{(-)}$.
Thus we have proved (b.i).

Let us next show that the state $\oip$ is disjoint from $\mvzi$
 on $\mAoi(\Oo')$
for every double cone $\Oo'$. We begin 
by noting that the unitaries $Z^{(n)}_w(r)$ are contained in the local
von Neumann algebras $\mAoi(\Oo_{n,w})^-$ for suitable double cones
$\Oo_{n,w}$
and, by local normality, the state $\oip$ extends to these
algebras.
On the other hand, we have that $Z^{(n+1)}_w(r) \in \mAoi(\Oo_{n,+})^-
\cup \mAoi(\Oo_{n,-})^-$ where $\Oo_{n,\pm}$ are double cones in the
right/left spacelike complement of the double cone  
 $\Oo_n$  based on the interval $(-na,na)$. This follows from
 the support properties of the functions $g_n$
and the Weyl-relations.  Since $Z_w^{(n)}(r)$ forms a central sequence
 in $\mAoi(\Oo')^-$, it is sufficient for
the proof of the claimed disjointness  to show
that, for each  $\epsilon > 0$, there is some 
real-valued
$w \in \DR$ with $w(\vx) = 1$ for $|\vx| < 1$, such that in the limit of
large $n$,
\begin{equation}
|\, \mvzi(Z^{(n)}_w(\pi/q)) - \oip(Z^{(n)}_w(\pi/q))\,| \ge 2 -
\epsilon\,.
\end{equation}
That this can be achieved may be shown as follows.
Let
${\cal C} := \{w \in \DR:w\ {\rm real},\  w(\vx) = 1
 \  {\rm for} \  |\vx| < 1\}$, and
  let $\epsilon >0$ be given.  We claim that there is some 
$w \in \cal C$ having the property
\begin{equation}
2\cdot {\rm e}^{-(\pi^2/4q^2)\cdot \int
 d\vk\,|\vk|\,|\tilde{w}(\vk)|^2} \ge 2 -
\epsilon\,.
\end{equation}
It is plain that this amounts to showing that
there holds
\begin{equation}
\inf_{w \in {\cal C}}\, \int d\vk\,|\vk|\,|\tilde{w}(\vk)|^2 \ = \ 0\,.
\end{equation}
Now $w \in \cal C$ implies that also $w_j \in \cal C$, where
\begin{equation}
w_j(\vx) := w(\vx/j)\,, \quad   j \in \NN\,. 
\end{equation}
On the other hand, $\int d\vk\, |\vk|\, |\widetilde{w_j}(\vk)|^2=
\int d\vk\, |\vk|\, |\tilde{w}(\vk)|^2 $
for all $j \in \NN$, and for all $u \in \DR$
there holds in the limit of large $j$
\begin{equation}
\int d\vk\,u(\vk)\,|\vk|^{1/2}\widetilde{w_j}(\vk)
 = \int d\vk\,j^{-1/2}u(j^{-1}\vk)\,|\vk|^{1/2}\tilde{w}(\vk)
\to 0 \,.
\end{equation}
This shows that the sequence of functions $\vk \mapsto
 |\vk|^{1/2}
\widetilde{w_j}(\vk)$,  $j \in \NN$, tends weakly to $0$ in $L^2(\RR)$
 for $j\to \infty$. Since $\cal C$
 is a convex set there
 exists then another, averaged
 sequence $\overline{w_j} \in \cal C$ such that $\int d\vk\,
 |\vk|\,|\tilde{\overline{w_j}} (\vk)|^2 \to 0$ for $j \to 0$.
 Thus we arrive at (4.36), and
 consequently there is for the prescribed $\epsilon > 0$ a $w \in
 \cal C$ satisfying (4.35).

Our next claim is that, with such a $w \in \cal C$,  relation (4.34)
is satisfied for all $n \in \NN$. To this end we recall that
in view of the Weyl-relations and locality there is for any
choice of $n \in \NN$  some $j \in \NN$ such that
\begin{eqnarray}
\oip(Z_w^{(n)}(\pi/q)) & = & \langle
V_q^{(j)}\Ooi,Z_w^{(n)}(\pi/q) V_q^{(j)}\Ooi \rangle \\
& = & {\rm e}^{i(\pi/q)\int d\vx\, u_j(\vx)g_n(\vx)}\langle
\Ooi,Z_w^{(n)}(\pi/q)\Ooi \rangle \,. \nonumber
\end{eqnarray}
 For the integral in the exponential we obtain for sufficiently large $j$
\begin{equation}
\int d\vx\, u_j(\vx)g_n(\vx) = \int d\vx\,u_j(\vx)\partial_{\vx}
 w(\vx/n)
 = -q\,.
\end{equation}
This is a consequence of the support properties of the functions
involved:
$\partial_{\vx} w(\vx/n)$ is supported on two disjoint regions close
to $\pm n$, and
$u_j$ vanishes on the region at $-n$ and equals $q$ on the
other one. Therefore, with the help of Lemma 4.3 and the preceding
estimates,  we arrive at
\begin{eqnarray}
 |\,\mvzi(Z_w^{(n)}(\pi/q)) - \oip(Z_w^{(n)}(\pi/q))\,| & = &
 (1 - {\rm e}^{-i\pi})\langle \Ooi,Z_w^{(n)}(\pi/q)\Ooi \rangle \\
& = & 2 \cdot {\rm e}^{-(\pi^2/4q^2)\cdot
 \int d\vk\,|\vk|\,|\tilde{w}(\vk)|^2} \ \ge \
 2 - \epsilon \,. \nonumber
\end{eqnarray}
 This completes
the proof of statement (b.ii) in the theorem.

We mention as an aside that by the same argument one sees that also the
states $\oip$ corresponding to different charge values $q$ are
disjoint on the algebras $\mAoi(\Oo')$. Hence the charge of these states
can be determined in the spacelike
complement of any double cone, which may be interpreted as a
manifestation of Gauss' law.

It remains to establish the last statement (b.iii) of the theorem
concerning the implementation of the translations in the
representations induced by $\oip$. We indicate here only the essential
steps in the argument and refrain from giving the necessary
computations as they are of a similar nature as in the preceding
steps.

We begin by noting that the representation $\rqi$  of $\mAoi$, which
is fixed by $\oip$, can be realized on the Hilbert space $\Hoi$
by setting 
\begin{equation}
  \rqi(\poi(\Au)) := \lim_n \, \Vqn{}^*\poi(\Au)\Vqn\,,
\end{equation}
with $\Vqn$ as in (4.28). This limit exists in the norm topology
because of the Weyl-relations and locality. We note in passing that
$\rqi$ is an automorphism of the $C^*$-algebra generated by the local
von Neumann algebras $\mAoi(\Oo)^-$.

The translations in the representation $\rqi$ are obtained by setting
\begin{equation}
U_{q,\iota}(x) := \lim_n\, \Vqn{}^* U_{0,\iota}(x) \Vqn \,. 
\end{equation}
Here $U_{0,\iota}(x)$ are the unitary translation operators in the
defining representation of the net $\mAoi$ on $\Hoi$. The somewhat
laborious task is to show that this limit exists 
 for a suitable choice of the functions $u_n$ in the
definition of $\Vqn$. In contrast to the preceding results where the
specific form of $u_n$ was, for $\vx > n\,a$, completely arbitrary  (cf.\
(4.29)), the proof that the translations can be represented in the
form (4.43) requires a proper choice of these functions in that
region. With these preparations one can show that the operator
$\Vqn{}^* U_{0,\iota}(x) \Vqn U_{0,\iota}(x)^*$ can be decomposed into a
product $Z^{(n)}_x \, \Gamma_x$ of Weyl-operators, where $\Gamma_x$
does not depend on $n$ and $Z^{(n)}_x$ is a central sequence whose
vacuum expectation value converges to 1, uniformly on compact sets in
$x$. Since each $Z^{(n)}_x$ is unitary, the uniform convergence of (4.43)
 in the strong-operator topology follows.

It is then clear that (4.43) defines a continuous unitary representation of
the translations on $\Hoi$ which satisfies the relativistic spectrum
condition (since $U_{0,\iota}$ does). Combining relation (4.42) with
(4.43) one also sees that the unitaries $U_{q,\iota}(x)$ implement the
translations $\maoi_x$ in the representation $\rqi$ and this completes
the proof of the theorem.
%%%%%%%%%%%%%%%%%%%%%%%%%%%%%
\section{Some open problems}
\setcounter{equation}{0}
%%%%%%%%%%%%%%%%%%%%%%%%%%%%%
The present investigation of the short distance properties of free field
theories has produced some interesting results which corroborate the
general ideas expounded in \cite{BV}. Yet in spite of the basic
simplicity of the underlying class of models there remain some
intriguing questions whose understanding seems to be of importance for
the treatment of less trivial examples, notably interacting theories.

First, there is the role of dimension. In our computation of the
scaling limit of free field theories in Sec.\ 2 we relied heavily on
the fact that the corresponding vacuum states are, in $s = 2$ and $3$
dimensions, locally normal with respect to each other for any value $m
\ge 0$ of the mass. In $s > 3$ dimensions this is no longer true
because of ultraviolet problems and this fact resembles the
situation which one expects to encounter in interacting theories in
physical spacetime. There the scaled vacuum states (corresponding to
different ``running'' values of the coupling constants and masses) are
most likely locally disjoint. Hence it would be
of interest to develop in the simpler case of free field theory in $s
>3$ dimensions techniques which allow one to compute the scaling
limit nets $(\mAoi,\maoi)$ without relying on local normality. It is
clear from our present arguments that also in these models the local
algebras $\mAoi(\Oo)$ and $\Aa^{(0)}(\Oo)$ have large sub-algebras in
common which consist of smoothed-out Weyl-operators. But this
information is not yet sufficient in order to clarify the relation
between the respective nets.

Because of similar reasons we neither have an explicit description of
the scaling limit theory in $s=1$ dimensions, nor do we know whether
this scaling limit is unique. If one restricts attention to the
sub-algebra of the scaling algebra which is generated by smoothed-out
Weyl-operators, then one can show that the resulting subnets of the
scaling limit theory are isomorphic to the nets generated by
smoothed-out Weyl-operators in massless free field theory, tensored
with an Abelian algebra \cite{Bu96}. It is an open problem whether
an analogous result holds if one starts from the full scaling algebra.

Within the realm of free field theory it would also be of interest to
determine the scaling limit of the  local net generated by a free massive
vector field, which resembles in certain respects the Higgs model.
It would be interesting to see whether in the free field case there exist in
the scaling limit physical states which carry a charge that is
screened at finite scales, similarly as in the Schwinger model.

The real challenge, however, is the short distance analysis of
interacting theories with the method of the scaling algebra. There the
simplest examples are the ${\cal P}(\phi)_2$-models which are known to be
asymptotically free (super-renormalizable) and to have vacuum states
which are locally normal with respect to the massive Fock
vacuum. Because of this close relationship to free field theory one
may hope that the present results will also be of use in the analysis
of these more interesting examples.

\noindent {\Large \bf Acknowledgment} \\[3 mm] 
We gratefully acknowledge financial support by the DFG (Deutsche 
Forschungsgemeinschaft).
%%%%%%%%%%%%%%%%%%%%%%%%%%%%%%%%%%%%%%%%%%%%%%%%%%%%%%%%%%%%%%%%%%%%%%%%%%%%%
\newcommand{\CMP}{{\it Commun. Math. Phys. }}
\newcommand{\LMP}{{\it Lett. Math. Phys. }}
\newcommand{\RMP}{{\it Rev. Math. Phys. }}
%%%%%%%%%%%%%%%%%%%%%%%%%%%%%%%%%%%%%%%%%%%%%%%%%%%%%%%%%%%%%%%%%%%%%%%%%%%%%%


\begin{thebibliography}{122}
\bibitem{BR} O. Bratteli and D.W. Robinson, {\it Operator algebras
    and quantum statistical mechanics}, Vol. 2, 2nd ed., Springer-Verlag, 
New York, 1996
\bibitem{Bu96a} D. Buchholz, ``Phase space properties of local
    observables and structure of scaling limits'', {\it
    Ann. Inst. H. Poincar\'e} {\bf A 64} (1996) 433
\bibitem{Bu96} D. Buchholz, ``Quarks, gluons, colour: Facts or
    fiction?'', {\it Nucl. Phys.} {\bf B469} (1996) 333
\bibitem{BDF}  D. Buchholz, C. D'Antoni, and K. Fredenhagen,
``On the universal structure of local algebras'', \CMP {\bf 111} (1987)
123
\bibitem{BF} D. Buchholz and K. Fredenhagen, 
``Locality and the structure of particle states'', \CMP
 {\bf 84} (1982) 1
\bibitem{BJu}  D. Buchholz and P. Jacobi, ``On the nuclearity
  condition for massless free fields'', \LMP
  {\bf 13} (1987) 313
\bibitem{BuLu} D. Buchholz and M. Lutz, in preparation
\bibitem{BV}   D. Buchholz and R. Verch, ``Scaling algebras and
renormalization group in algebraic quantum field theory'',
  \RMP {\bf 7} (1995) 1195
\bibitem{EF} J.P. Eckmann and J. Fr\"ohlich,
``Unitary equivalence of local algebras in the quasi-free representation'' {\it
    Ann. Inst. H. Poincar\'e} {\bf A 20} (1974) 201
\bibitem{Ha} R. Haag, {\it Local quantum physics}, 2nd ed.,
    Springer-Verlag,
Berlin-Heidelberg-New York, 1996
%% \bibitem{HSw} R. Haag and J.A. Swieca, ``When does a quantum field
%%  theory
%% describe particles?'', \CMP {\bf 1} (1965) 308
\bibitem{LoSw} J.H. Lowenstein and J.A. Swieca, ``Quantum
  electrodynamics in two dimensions'',
{\it Ann. Phys. (N.Y.)} {\bf 68} (1971) 172
\bibitem{Rob1} J.E. Roberts, ``Some applications of dilatation
  invariance to structural questions in the theory of local
  observables'',
\CMP {\bf 37} (1974) 273
\bibitem{Wigh} A.S. Wightman, ``La th\'eorie quantique locale et la
    th\'eorie quantique des champs'', {\it
    Ann. Inst. H. Poincar\'e} {\bf A 1} (1964) 403
\end{thebibliography}
\end{document}